\documentclass[10pt,compsoc]{IEEEtran}
\usepackage[labelfont=bf,textfont={bf}]{caption}
\usepackage{alltt}
\usepackage{multirow}
\usepackage{graphicx}
\usepackage{color}

\usepackage{subcaption}
\usepackage{pifont}
\usepackage{boxedminipage}
\usepackage{notoccite}
\usepackage{array}
\usepackage{rotating}
\usepackage{graphicx}
\newcolumntype{C}[1]{>{\centering\let\newline\\\arraybackslash\hspace{0pt}}m{#1}}
\usepackage{enumerate}
\usepackage{booktabs}

\usepackage[framemethod=tikz]{mdframed}
\usepackage{lipsum}
\usetikzlibrary{shadows}
\usepackage{mathtools}
 \usepackage{paralist}
 \definecolor{light-gray}{gray}{0.80}
\usepackage{graphics}
\usepackage[shortlabels]{enumitem}
\usepackage{dblfloatfix}
\usepackage[utf8]{inputenc}
\usepackage{multicol}
\usepackage{graphicx}
\usepackage{wrapfig}
\usepackage{colortbl}
\usepackage[ruled,vlined]{algorithm2e}
\usepackage[final]{listings}
\usepackage[framemethod=tikz]{mdframed}

\usepackage{lipsum}
\usetikzlibrary{shadows}
\usepackage{mathtools}
 \usepackage{paralist}
 \definecolor{light-gray}{gray}{0.80}
\usepackage{graphics}
\usepackage[shortlabels]{enumitem}
\usepackage{dblfloatfix}
\newmdenv[
  tikzsetting= {fill=light-gray},
  linewidth=1pt,
  roundcorner=0pt, 
  shadow=false
]{myshadowbox}
\usepackage{colortbl} 
\usepackage{blindtext, graphicx}
\usepackage{textcomp}
\usepackage[hidelinks]{hyperref}
\usepackage{amsmath}
\usepackage{amsfonts}
\usepackage{caption}
\usepackage{color}
\usepackage[final]{listings}
\usepackage{balance}
\usepackage{picture}
\usepackage{relsize}
\usepackage{multicol}
\usepackage{soul}
\usepackage{enumitem}
\setitemize{noitemsep,topsep=0pt,parsep=0pt,partopsep=0pt,leftmargin=*}
\setenumerate{noitemsep,topsep=0pt,parsep=0pt,partopsep=0pt,leftmargin=*}
\usepackage{tikz}
\newcommand*\circled[1]{\tikz[baseline=(char.base)]{ \node[shape=circle,draw,inner sep=0.1pt] (char) {#1};}}
\usepackage{makecell}
\usepackage{bigstrut}

\definecolor{comment_color}{rgb}{0.5, 0, 1}
\definecolor{steel}{rgb}{0.1, 0.3, 0.5} 

\newcommand{\bi}{\begin{itemize}}
\newcommand{\ei}{\end{itemize}}

\newcommand{\BLACK}{\color{black}}

\usepackage[
 pass,
]{geometry}
\usepackage[skins]{tcolorbox}
\usepackage{verbatim}
\usepackage{algorithmicx}
\usepackage{algpseudocode}
\usepackage[export]{adjustbox}
\renewcommand{\footnotesize}{\scriptsize}

\definecolor{black}{rgb}{0,0,0}
\definecolor{LightCyan}{rgb}{0.88,1,1}
\definecolor{darkgray}{gray}{0.7}
\definecolor{Gray}{rgb}{0.88,1,1}
\definecolor{Gray}{gray}{0.85}
\definecolor{Blue}{RGB}{0,29,193}
\definecolor{MyDarkBlue}{rgb}{0,0.08,0.45} 
\definecolor{pink}{rgb}{231,95,110}
\definecolor{lightergray}{rgb}{0.85, 0.85, 0.85}
\definecolor{darkgray}{rgb}{0.47, 0.47, 0.47}
\definecolor{lightestgray}{rgb}{0.95, 0.95, 0.95}
\definecolor{ao(english)}{rgb}{0.0, 0.5, 0.0}
\usepackage[tikz]{bclogo}
{\noindent\begin{minipage}[c]{\linewidth}%
\begin{bclogo}[couleur=gray!20,%
                arrondi=0,logo=\none,%
                ombre=true%
                ]{{\small  ~#1}}}%
{\end{bclogo}\vspace{2mm}\end{minipage}}

\usepackage{fancyvrb}
\fvset{%
fontsize=\small,
numbers=left
}
\newcommand{\be}{\begin{enumerate}}
\newcommand{\ee}{\end{enumerate}}

\definecolor{lightgray}{gray}{0.7}
\tikzstyle{highlighter} = [lightgray, line width = \baselineskip]
\usepackage{wrapfig}
\newcounter{highlight}[page]

\AtBeginShipout{\AtBeginShipoutUpperLeft{\ifthenelse{\value{highlight} > 0}{\tikz[remember picture, overlay]{\foreach \stroke in {1,...,\arabic{highlight}} \draw[highlighter] (begin highlight \stroke) -- (end highlight \stroke);}}{}}}

\newcommand{\squishlist}{
 \begin{list}{$\bullet$}
 { \setlength{\itemsep}{0pt}
   \setlength{\parsep}{3pt}
   \setlength{\topsep}{3pt}
   \setlength{\partopsep}{0pt}
   \setlength{\leftmargin}{1.5em}
   \setlength{\labelwidth}{1em}
   \setlength{\labelsep}{0.5em} } }

\newcommand{\squishlisttwo}{
 \begin{list}{$\bullet$}
 { \setlength{\itemsep}{0pt}
  \setlength{\parsep}{0pt}
  \setlength{\topsep}{0pt}
  \setlength{\partopsep}{0pt}
  \setlength{\leftmargin}{1em}
  \setlength{\labelwidth}{1.5em}
  \setlength{\labelsep}{0.5em} } }

\newcommand{\squishend}{
 \end{list} }

\usepackage{color}

\definecolor{awesome}{rgb}{1.0, 0.13, 0.32}

\definecolor{Gray}{gray}{0.95}

\usepackage{cite}

\usepackage{tcolorbox}
\newtcolorbox{blockquote}{colback=gray!5,boxrule=0.4pt,colframe=black,fonttitle=\bfseries}

\begin{document}

\title{     Defect Reduction Planning (using TimeLIME)     }

\author{Kewen~Peng, 
        Tim~Menzies,~\IEEEmembership{Fellow,~IEEE}
\IEEEcompsocitemizethanks{\IEEEcompsocthanksitem K. Peng and  T. Menzies are with the Department
of Computer Science, North Carolina State University, Raleigh, USA.
 \protect
E-mail:kpeng@ncsu.edu,   timm@ieee.org}}

\markboth{IEEE Transactions on Software Engineering}%
{Peng \MakeLowercase{\textit{et al.}}: TimeLIME defect reduction for IEEE Journals}

\IEEEtitleabstractindextext{
\begin{abstract}
Software comes in releases. An implausible change to software is something that has never been changed in prior releases. When planning how to  reduce defects, it is  better  to  use  plausible  changes,  i.e.,   changes  with some precedence in the prior releases.

To demonstrate these points, this paper compares several defect reduction planning tools. LIME is a local sensitivity analysis tool that can report the fewest changes needed to alter the classification of some code module (e.g.,  from ``defective'' to ``non-defective''). TimeLIME is a new tool, introduced in this paper, that improves LIME by restricting its plans to just those attributes which change the most within a project.   

In this study, we compared the performance of LIME and TimeLIME and several  other defect reduction planning algorithms. The generated plans were assessed via (a) the similarity scores between the proposed code changes and the real code changes made by developers; and (b) the   improvement scores seen within projects that followed the plans. For nine project trails, we found that TimeLIME outperformed all other algorithms (in 8 out of 9 trials). Hence, we strongly recommend using past releases as a source of knowledge for computing fixes for new releases (using TimeLIME).

Apart from these specific results, the other lesson from  this paper is that our community might be more careful about using off-the-shelf AI tools, without first applying SE knowledge (e.g. that past releases are a good source of knowledge for planning defect reductions). As shown here, once that SE knowledge is applied, this can result in dramatically better reasoning.
\end{abstract}

}

\maketitle
\IEEEpeerreviewmaketitle
\IEEEdisplaynontitleabstractindextext


\section{Introduction}


\begin{raggedleft}
{\em ``Don’t tell me where I am, tell me where to go.''\\
-- a (very busy) developer\\}
\end{raggedleft}
Machine learners generate models. People read models. What learners generate the kind of models that people want to read?
If the reader is a busy software developer, then they might not need, or be able to use, complex models. Rather, such a busy developer might instead just want to know the {\em least} they need to do to achieve the {\em most} benefits.  
For example,
 suppose some AI model has classified a module as ``defective''. If  a developer then asks ``what can I do to fix that?'' then, ideally, we should, be able to  reflect on the model  to learn a  defect reduction plan; i.e.,  a small set of  actions that reduces the odds of that module being defective.  
 
 For many machine learning algorithms, it can be (very)
difficult  to learn a succinct reduction plan by  reflecting on the arcane internal structure of, say, a neural net classifier.
To better support busy developers,
explanation algorithms like  LIME~\cite{ribeiro2016should} (first presented at KDD’16)  can report what attribute
 changes
can alter a classification (e.g.,  from ``defective'' to ``non-defective'').  But classic LIME has a problem-
it   generates surprising and unprecedented plans that had never been seen before in the history of a project.  As we show,
such unprecedented plans are sub-optimal.

To fix this problem, TimeLIME
adds SE knowledge to LIME.
We note that
software comes in releases
and that  
an implausible change to software is something that has never been changed in prior releases.  Hence, we propose 
the following {\em TimeLIME tactic}:
\begin{quote}
{\em 
When reasoning about changes
to a project, it is best
to use changes seen in the historical record of that project.}
\end{quote}
\noindent
To assess the value of this TimeLIME tactic,
we ask:
\bi
\item
{\bf RQ1: Does TimeLIME provide succinct plans?}
Classic LIME,   proposes changes to   dozens of attributes.
     TimeLIME, on the other hand, restricts itself to just the most changed attributes. Hence,
 our plans
     are easier to apply.
 \item
 {\bf RQ2: Could developers apply the changes proposed by TimeLIME?} 
Given project information divided into 
{\em oldest}, {\em  newer}, and {\em most recent} data,   this paper:
\bi
\item Used the {\em oldest} data to determine what attributes were often changed in a project,
\item Used the {\em newer} data to build plans using LIME, TimeLIME, and five other planning algorithms;
\item  Divided the {\em most recent}  data into:
\bi
\item
Those projects that {\em followed} the plans;
\item And those that did not. This study found a  large overlap (median=80\%)  between 
 TimeLIME's recommendations and  actual actions made by developers.
 \ei
 \ei
 \item
  {\bf RQ3: Is TimeLIME better at defect reduction?}
TimeLIME's plans  perform best (compared
to classic LIME and four other algorithms).
     \ei 
 \noindent
The rest of this paper is structured as follows.
\S\ref{background} discusses defect prediction, code refactoring, and challenges of using human opinions in SE.\S\ref{prior} introduces some prior works in the field of defect reduction and their methodologies. \S\ref{timelime} presents the basic framework of LIME as well as TimeLIME, the new method proposed in this paper.   \S\ref{K-test} shows our method for ranking different planning
methods. \S\ref{experiment} describes  experiment and  the  datasets, predictive model, and planners evaluated in this work.
\S\ref{result} and \S\ref{discussion} report and discuss the result respectively. The credibility and reliability of our conclusions is discussed by \S\ref{threat}. Recent related works are shown in \S\ref{related}, which also declares the major difference that distinguishes the contribution of this paper. Future work and directions are illustrated in \S\ref{futurework}. Finally, we conclude this work in \S\ref{conclusion}. 

\subsection{Reproduction Package}

All our scripts and data  are available on-line\footnote{https://github.com/ai-se/TimeLIME}.

\begin{table*}[t!]
\centering
\scriptsize
\begin{tabular}{l|l|l}
\hline
\rowcolor[HTML]{C0C0C0} 
Metric  & Name                        & Description                                                                                                     \\ \hline
amc     & average method complexity   & Number of JAVA byte codes                                                                                       \\ \hline
avg\_cc & average McCabe Average      & McCabe’s cyclomatic complexity seen in class                                                                    \\ \hline
ca      & afferent couplings          & How many other classes use the specific class.                                                                  \\ \hline
cam &
  cohesion amongst classes &
  \begin{tabular}[c]{@{}l@{}}Summation of number of different types of method parameters in every method divided by a multiplication \\ of number of different method parameter types in whole class and number of methods.\end{tabular} \\ \hline
cbm     & coupling between methods    & Total number of new/redefined methods to which all the inherited methods are coupled                            \\ \hline
cbo     & coupling between objects    & Increased when the methods of one class access services of another.                                             \\ \hline
ce      & efferent couplings          & How many other classes is used by the specific class.                                                           \\ \hline
dam     & data access                 & Ratio of private (protected) attributes to total attributes                                                     \\ \hline
dit     & depth of inheritance tree   & It’s defined as the maximum length from the node to the root of the tree                                        \\ \hline
ic      & inheritance coupling        & Number of parent classes to which a given class is coupled (includes counts of methods and variables inherited) \\ \hline
lcom    & lack of cohesion in methods & Number of pairs of methods that do not share a reference to an instance variable.                               \\ \hline
lcom3 &
  another lack of cohesion measure &
  \begin{tabular}[c]{@{}l@{}}If $m$, $a$ are the number of methods, attributes in a class number and $\mu(a)$ is the number \\ of methods accessing an attribute, then lcom3 = $((\frac{1}{a}\sum_{j}^{a} {\mu(a_j)})-m)/(1-m)$ \end{tabular} \\ \hline
loc     & lines of code               & Total lines of code in this file or package.                                                                    \\ \hline
max\_cc & Maximum McCabe              & Maximum McCabe’s cyclomatic complexity seen in class                                                            \\ \hline
mfa     & functional abstraction      & Number of methods inherited by a class plus number of methods accessible by member methods of the class         \\ \hline
moa     & aggregation                 & Count of the number of data declarations (class fields) whose types are user defined classes                    \\ \hline
noc     & number of children          & Number of direct descendants (subclasses) for each class                                                        \\ \hline
npm     & number of public methods    & Npm metric simply counts all the methods in a class that are declared as public.                                \\ \hline
rfc     & response for a class        & Number of methods invoked in response to a message to the object.                                               \\ \hline
wmc & weighted methods per class &
  A class with more member functions than its peers is considered to be more complex and more error prone. \\ \hline
defect  & defect                      & Number of bugs which can be transformed into Boolean values for classification.   \\ \hline                                          
\end{tabular}

\vspace{5mm}
\caption{The C-K OO metrics used in defect prediction. The last variable "defect" is the dependent variable.}
\label{ck}
\end{table*}

\section{Background }\label{background}
\subsection{Challenges with Using Human Opinions}

This paper is an {\em algorithmic analysis of historical SE data}
where we  ran simulations over the historical record of eight software projects.
An alternate approach to this algorithmic analysis of historical SE data is to    use {\em qualitative methods}. Qualitative methods  rely on surveys of human subject matter experts (e.g.,  programmers). Much has been learned from such studies of subject matter experts~\cite{begel2013social}. Nevertheless, in the particular case of large scale defect prediction, we prefer our algorithmic approach, for two reasons:

\bi
\item \textit{Scalability}: It is hard to scale qualitative  investigations of human beliefs   to a large number of projects. 
We mention this since while this paper studies just
eight projects, our long-term goal is to develop software analysis methods that applies to hundreds to thousands of projects.
While some progress has been seen recently with scaling qualitative methods~\cite{chen2019replication}, at the time of this writing, we  assert that it is far easier to scale an algorithmic analysis of historical SE data.

\item \textit{Lack of consensus}: multiple studies report that human beliefs in software quality may often be inconsistent and even incorrect. Devanbu et al. have conducted a case study among 564 Microsoft software developers to show that human beliefs on software quality can be quite varied and may not necessarily correspond with actual evidence within current projects~\cite{devanbu2016belief,bird2011don}. Similar assertions are also made in Passos' paper, where the author reports that conflicting beliefs can be held by different stakeholders of the software development team. There also exist cases that a belief is correct for past projects but not the current work~\cite{passos2011analyzing}). 
A more recent study by Shrikanth et al. also reports such much variability of human beliefs about defect prediction~\cite{shrikanth2019assessing}. Shrikanth studies 10 beliefs held by software developers about defect prediction, which were initially summarized by Wan et al in 2018  \cite{wan2018perceptions}. By measuring the actual support of these beliefs within the project, Shrikanth found that:
\bi[label=$\circ$]
\item
Among over 300,000 changes seen in different open-source projects, only 24\% of the projects support all 10 beliefs.
\item
What is believed the most by developers does not necessarily have the strongest support within projects.
For example, a belief that is acknowledged by 35\% of the developers has the most support whereas a belief held by 76\% of the developers is only ranked 7th out of 10 beliefs.
\item
As a project grows to mature, the beliefs actually tend to be weakened rather than strengthened.
\ei
Not only do practitioners have conflicting beliefs about what causes defects, but we also can see that researchers who have studied many projects also disagree on what factors matter the most to defect reduction. For example, as discussed later in  the paper, Alves~\cite{alves2010deriving}, Shatnawi~\cite{shatnawi10g1}, and Oliveira~\cite{oliveira2014extracting} all offer different models about what matters most for software quality.  
\ei
In summary, many studies report a significant disconnect between human beliefs and patterns supported by data. Hence, we
are nervous about using  the opinion of experts' opinions as the ``ground truth'' to evaluate (e.g., ) defect
reduction plans. Accordingly, we use an algorithm analysis
since that can use   historical SE data to generate the ground truth needed to evaluate  a method.


\begin{table*}[!b]
\centering
\scriptsize
\begin{adjustbox}{max width=\textwidth}
\begin{tabular}{|l|c|c|c|c|c|c|c|c|c|c|c|c|c|c|c|c|c|c|c|c|}
\hline
\multicolumn{1}{|l|}{} &
  \multicolumn{1}{l|}{\rotatebox{90}{AMC}} &
  \multicolumn{1}{l|}{\rotatebox{90}{AVG\_CC}} &
  \multicolumn{1}{l|}{\rotatebox{90}{CA}} &
  \multicolumn{1}{l|}{\rotatebox{90}{CAM}} &
  \multicolumn{1}{l|}{\rotatebox{90}{CBM}} &
  \multicolumn{1}{l|}{\rotatebox{90}{CBO}} &
  \multicolumn{1}{l|}{\rotatebox{90}{CE}} &
  \multicolumn{1}{l|}{\rotatebox{90}{DAM}} &
  \multicolumn{1}{l|}{\rotatebox{90}{DIT}} &
  \multicolumn{1}{l|}{\rotatebox{90}{IC}} &
  \multicolumn{1}{l|}{\rotatebox{90}{LCOM}} &
  \multicolumn{1}{l|}{\rotatebox{90}{LCOM3}} &
  \multicolumn{1}{l|}{\rotatebox{90}{LOC}} &
  \multicolumn{1}{l|}{\rotatebox{90}{MAX\_CC}} &
  \multicolumn{1}{l|}{\rotatebox{90}{MFA}} &
  \multicolumn{1}{l|}{\rotatebox{90}{MOA}} &
  \multicolumn{1}{l|}{\rotatebox{90}{NOC}} &
  \multicolumn{1}{l|}{\rotatebox{90}{NPM}} &
  \multicolumn{1}{l|}{\rotatebox{90}{RFC}} &
  \multicolumn{1}{l|}{\rotatebox{90}{WMC}} \\ \hline
\circled{1} inline methods   & dec? & inc  &     &      &  &     &     &  &     &     &      &      & dec? & inc? &  &      &     & dec? & dec & dec \\ \hline
\circled{2} extract method   & inc? & dec  &     &      &  &     &     &  &     &     &      &      & inc? & dec? &  &      &     & inc? & inc & inc \\ \hline
\circled{3} extract class    & dec  & dec  &     & inc? &  & inc & inc &  &     &     & dec? & dec? & dec  & dec? &  & dec? &     &      & dec & dec \\ \hline
\circled{4} inline class     & inc  & inc  &     & dec? &  & dec & dec &  &     &     & inc? & inc? & inc  & inc? &  & inc? &     &      & inc & inc \\ \hline
\circled{5} move method      & dec  & inc? &     &      &  &     &     &  &     &     &      &      & dec  & dec? &  & dec? &     & dec? &     & dec \\ \hline
\circled{6} hide delegate    &      &      & dec &      &  & dec &     &  &     &     &      &      &      &      &  &      &     &      &     &     \\ \hline
\circled{7} consolidate cond & dec  & dec  &     & dec  &  &     &     &  &     &     & inc  & inc  & dec  & dec? &  &      &     &      & inc & inc \\ \hline
\circled{8} polymorphism     & dec  & dec  &     &      &  &     & inc &  & inc & inc &      &      &      & dec? &  &      & inc &      &     &     \\ \hline
\circled{9} flatten cond      & dec & dec  &     &      &     &  &     &  &  &  &      &      & dec & dec? &     &  &     &  &     &     \\ \hline
\circled{10} hide method       &     &      &     & inc  &     &  &     & inc  &  &  &   &   &     &      &     &  &     &  &     &     \\ \hline
\circled{11} simplify para     & dec & dec  &     &      &     &  &     &  &  &  &      &      &     & dec  &     &  &     &  &     &     \\ \hline
\circled{12} factory method    & inc & inc  &     &      &     &  &     &  &  &  &      &      & inc & inc? &     &  &     &  & inc & inc \\ \hline
\circled{13} push down method  & dec & dec? &     &      & dec &  &     &  &  &  &      &      &     &      &     &  &     &  & dec & dec \\ \hline
\circled{14} encapsulate field &     &      &     & dec  &     &  &     &  &  &  & inc  & inc  & inc &      &     &  &     &  &     & inc \\ \hline
\circled{15} extract subclass  & dec & dec  & inc & inc? &     &  & inc &  &  &  & dec? & dec? & dec &  dec?    & dec &  & inc &  & dec & dec \\ \hline
\circled{16} inline subclass   & inc & inc  & dec & dec? &     &  & dec &  &  &  & inc? & inc? & inc &   inc?   & inc &  & dec &  & inc & inc \\ \hline
\end{tabular}
\end{adjustbox}
\caption{
Code refactoring methods.
Taken from~\cite{Shvets:2014}.
In this table ``dec`` and ``inc'' are short hand for decrease and increase (respectively). These cell values were determined  as follows. For each of the methods in column one, ten times,
we applied that refactoring method to some randomly selected portion of the code base used in this study. The measurements listed in the columns headers were collected before and after that change.
 Empty cells indicate no influence on the feature. Cells marked with "?" means we observed that some examples have the change while others do not.
 }
\label{tab:refactoring}
\end{table*}

\subsection{Defect Prediction}\label{defect}

The case study of this paper comes from defect prediction and planning. This section
discussed the value of that kind of analysis.

During software development,   testing    often has some resource limitations.
For example, the effort associated with coordinated human effort across a large code base can   grow exponentially with the scale of the project \cite{fu2016tuning}.
Hence, to effectively manage resources, it is common to match the quality assurance (QA) effort to the perceived criticality and bugginess of the code.
Since every decision is associated with a human and resource cost to the developer team, it is impractical and inefficient to distribute equal effort to every component in a software system\cite{briand1993developing}.   Learning  defect prediction (using data miners) from static code attributes (like those shown in Table \ref{ck}) is one very cheap way to ``peek'' at the code and decide where to spend more QA effort.

Recent results show that software defect predictors are also competitive widely-used  automatic methods.  
Rahman et al.~\cite{rahman2014comparing} compared (a) static code analysis tools FindBugs, Jlint, and PMD with (b) defect predictors (which they called ``statistical defect prediction'') built using logistic  regression.
No significant differences in   cost-effectiveness were observed.
Given this equivalence, it is significant to
note that  defect prediction can be quickly adapted to new languages by building lightweight parsers to extract  code metrics. The same is not true for static code analyzers - these need extensive modification before they can be used in new languages.
Because of this ease of use,  and its applicability to many programming languages, defect prediction has been   extended  many ways including:
\begin{enumerate}[leftmargin=16pt,topsep=4pt]
\item Application of defect prediction methods to locating code with security vulnerabilities~\cite{Shin2013}.
    \item Understanding the factors that lead to a greater likelihood of defects such as defect prone software components using code metrics (e.g.,  ratio comment to code, cyclomatic complexity) \cite{menzies10dp, menzies07dp} or process metrics (e.g.,  recent activity).
    \item Predicting the location of defects so that appropriate resources may be allocated (e.g., ~\cite{bird09reliabity})
    \item Using predictors to proactively fix  defects~\cite{arcuri2011practical}
    \item Studying defect prediction not only just release-level \cite{chen2018applications} but also change-level or just-in-time \cite{commitguru}.  
    \item Exploring ``transfer learning'' where predictors from one project are applied to another~\cite{krishna2018bellwethers,nam18tse}.
    \item Assessing different learning methods for building predictors~\cite{ghotra15}. This has led to the development of hyper-parameter optimization and better data harvesting tools \cite{agrawal2018wrong, agrawal2018better}. 
\end{enumerate}
\newpage This paper extends defect prediction and planning in yet another way: exploring the  trade-offs between explanation and planning and the performance of defect prediction models. But beyond the specific scope of this paper, there is nothing in theory stopping the application of this paper to all of the seven areas listed above (and this would be a fruitful area for future research).

\subsection{Code refactoring}\label{refactoring}

Code refactoring is an important part of  software maintenance. The process is meant to improve the internal quality of software by better structuring the existing code, without changing the external behavior~\cite{fowler2018refactoring,wake2004refactoring}. Such restructuring is assumed to positively affect the software quality by reducing complexity, enhancing maintainability, etc.~\cite{mens2004survey,du2004refactoring}.

Much research has studied the relation between the code refactoring process and  internal software static attributes
metric and external software quality attributes like maintainability, modifiability, and understandability
~\cite{alshayeb2009empirical,geppert2005refactoring,moser2006does,simon2001metrics}. 
 Studies have  shown a correlation between external quality attributes and internal quality attributes (such as the OO metrics used in this paper)~\cite{dandashi2002method,bruntink2006empirical,tahvildari2004quality,wilking2007empirical,du2005does}.

That said, a missing piece of current research is what we call the planning
problem. Given that a developer {\em can} change these metrics
in many ways, what should she {\em actually do}?
How do we bridge the gap between ``what'' to refactor and ``why'' we need to refactor in the first place?
Table~\ref{tab:refactoring} shows the effect of various code refactoring methods on  our code base.  In order to select ``what'' refactoring method to apply, developers need some mapping from those refactoring to some higher-level goal. The planning algorithms of this paper provide that mapping to the higher-level goal of ``defect reduction in future releases''.

\BLACK 

\section{Prior Work in Planning  Defect\newline Reduction}\label{prior}

Over the years, several researchers have proposed various ways to identify appropriate changes on code metrics. This section will illustrate 4 methods that rely on either {\em outlier statistics} or {\em cluster deltas}. 

{\em Outlier statistics:} The general principle underlying outlier statistics methods is that in the distribution of values for each code metric, there are some extremely large/small values that are associated with greater defect proneness. Therefore, by changing those metrics to not have such outlier values, the code base may be found fewer bugs. This paper presents 3 outlier statistics methods and the major distinction among them is their different ways to identify the threshold for outlier values.
In the following text, the methods of Alves et al., Oliveira et al, Shatnawi are based on  outlier statistics. 

{\em Cluster deltas} is a framework for learning conjunctions of rules that need to be applied to the code metrics simultaneously. Unlike outlier statistics, which merely studies the statistical distribution of code metrics, cluster deltas is a supervised learner that take account of whether the code base is defective.
In the following text, Krishna's XTREE method uses cluster deltas to learn association rules concerning about when and where to apply a code change. 
 
These two approaches are discussed below. Before doing that, we first digress to make the point that {\em none} of the following
 methods  can be effective unless:
 \be
 \item It can be shown that programmers
 can apply the suggestions made in these
 plans;
 \item It can also be shown that when
 programmers apply the suggestions,
 they do not inadvertently add other
 changes that reduce (or remove)
 the  effectiveness of these plans.
  \ee
Later in this paper, we show
evidence that these points 1,2
are actually achievable -- see \S\ref{discussion}. 
\subsection{Alves. 2010}
Alves et al.~\cite{alves2010deriving} offers an unsupervised approach that learns from the statistical distribution and scale of OO metrics. At the beginning, Alves' method will weight each metric value according to the lines of code (LOC in Table~\ref{ck}) of its code class. The weighted metric values will then be normalized by the total sum of weights and sorted in an ascending order. Note that the sorted result is just equivalent to a cumulative probability function where x-axis stands for the weight percentage from 0 to 100\% and y-axis the metric scale. 

After that, a threshold percentage will be customized (Alves et al. recommends 70\%) to identify normal metric values against abnormal metric values. For example, a threshold of 70\% will identify the value for each metric where 70\% of the classes fall below. The intuition behind this is straightforward: they believe that a code class with outlier metric values that exceed 70\% of its peers is more likely to be found bugs. 

When we implemented the Alves' method in our experiment, we augmented the original implementation by also studying the correlation between the code metrics and the defect state of the class. By fitting each dependent variable and the independent variable with a univariate logistic regression classifier:
\bi
\item
we were able to reject metrics that
are poor indicators of defects (here we define "poor" as a logistic regression with $p$-value $>0.05$). 
\item
For those metrics that survived from the rejection, the planner will identify the normal range according to the threshold, i.e.,  [0, 70\%] for each metric. 
\item
Finally, during the planning process, any "survived" metric exceeding the threshold value will be proposed to reduce its value to the normal range.
\ei
\subsection{Shatnawi, 2010}
Shatnawi~\cite{shatnawi10g1} in 2010 provided an alternative to Alves' method by using VARL (Value of Acceptable Risk Level) to compute the outlier threshold. Initially proposed by Bender~\cite{bender1999quantitative} in 1999 in his epidemiological studies, the VARL function is a supervised learner that uses the interpretation of the univariate logistic regression model to derive the threshold for an acceptable risk level given by a probability $p_0$ (i.e.,  $p_0=0.05$). That is to say, the VARL believes that the probability $p_0$ of an event is less than 0.05 of the value of the dependent variable is smaller than VARL. The VARL function is as follow:
\begin{quote}
\centering
    $VARL = \frac{1}{\beta} (log(\frac{p_0}{1-p_0})-\alpha)$
\end{quote}
Here,$\alpha$ is the intercept of the logistic regression, $\beta$ is the coefficient of the logistic regression, and $p_0$ is the acceptable risk probability as described above. 

Similar to our procedure of implementing Alves' method, we ruled out metrics with $p$-value $>0.05$, and computed the VARL for the remaining metrics. We define the proposed plan for each metric as $[0, VARL]$, which means a metric value exceeding VARL will be recommended a reduction by the planner.
\begin{table*}[!b]
~\hrule~\vspace{0.1em}
\begin{minipage}{.49\linewidth}
\includegraphics[height = 4.5cm,width=.9\linewidth]{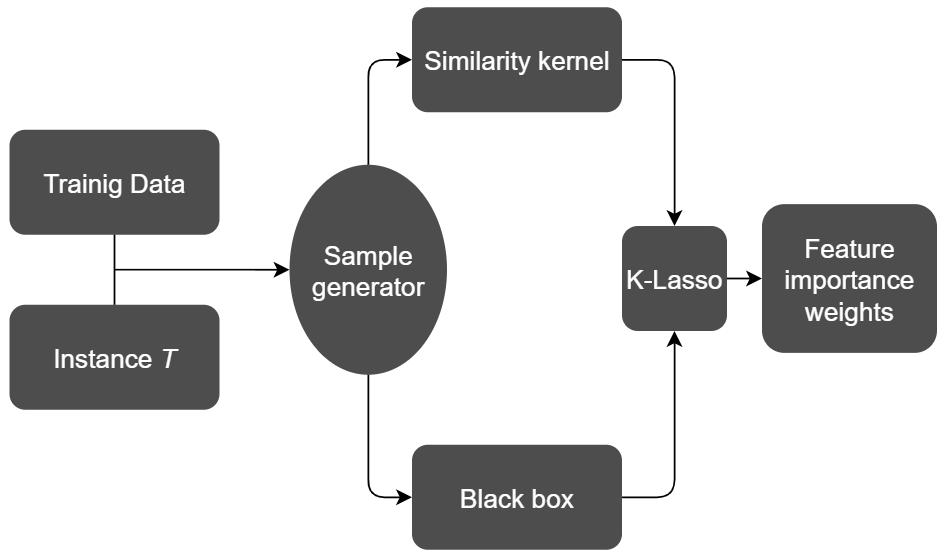}
\end{minipage}
\begin{minipage}{.49\linewidth}
\small
\begin{itemize}
\item
LIME is designed to be an add-on to other AI systems
(e.g.,  neural network, support vector machine, and so on).
Hence, it treats those AI tools as a 
``\emph{black box}'' that is queried within its processing.
\item
Within LIME, some  \emph{sample generator} is used to generate synthetic data which later gets passed to  the  \emph{black box}  and a \emph{similarity kernel}, along with the original \emph{training data}. 
\item
The \emph{similarity kernel} is an instrument used to weight the prediction results of \emph{training data} returned by the \emph{black box} by how similar they are to the \emph{instance T}. 
\item
The \emph{K-Lasso} is the procedure that learns the importance weights from the $K$ features selected with Lasso using a class of linear models.
\vspace{2mm}
\end{itemize}
\end{minipage}
~\hrule~
\caption{Inside LIME. From~\cite{ribeiro2016should}. The feature importance weights are passed to Algorithm \ref{simpleflip} and \ref{refineflip}, as later elaborated in \S\ref{planner}.
For a sample of the output feature importance
weights, see Figure~\ref{fig:lime}.}\label{inside}
\end{table*}

\subsection{Oliveira, 2014}
Oliveira et al.~\cite{oliveira2014extracting} approach an totally different threshold definition than the previous 2 methods. Instead of deriving an absolute threshold like Alves et al. and Shatnawi did, Oliveira et al. choose to use the \emph{relative threshold}, which indicates the percentage of classes the the upper bound (threshold) shall be applied to. The general format of their defect reduction rules is as follow:
\begin{quote}
\centering
    {\em p\% of the classes must have $M \leq K$}
\end{quote}
Here, $M$ is the code metric; $K$ is the threshold value for the corresponding metric; $p\%$ is the minimum percentage of code classed that are required to follow the restriction specified above.  

In order to compute the pair of values $(p,K)$ for each metric $M$, Oliveira defines 3 functions: \textbf{Compliance($p,k$)}, \textbf{Penalty1($p,k$)}, and \textbf{Panelty2($p,k$)}. The \textbf{Compliance} method reports the percentage of classes that follow the rule defined by each pair of values $(p,K)$. The \textbf{Penalty1} penalizes the model if the compliance rate is lower than a constant percentage (i.e.,  90\%). \textbf{Penalty2} computes the distance between $k$ and the median of the preset $Tail$-th percentile for each metric (Oliveira et al. suggest $90$-th percentile). Summing up the 2 penalty values to obtain the total penalty, the method chooses the pair of values $(p,K)$ with the lowest total penalty where a tie will be broken by choosing the highest $p$ and the lowest $k$. 

\definecolor{aoenglish}{rgb}{0.0, 0.5, 0.0}
 \definecolor{applegreen}{rgb}{0.74, 0.85, 0.75}

\begin{figure}[t!]
    \includegraphics[width=\linewidth]{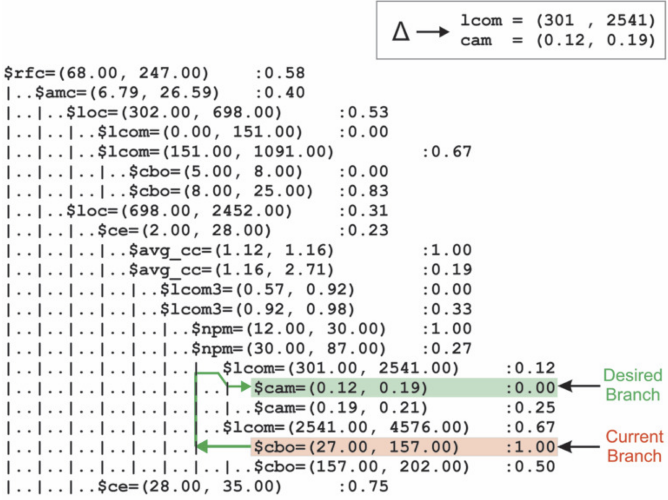}
    \caption{ XTREE plan generation. From Krishna et al.~\cite{krishna2017learning}. 
    An example has fallen down to
    the \textcolor{orange}{{\bf current branch}} where the probability of defects is 1.00.   The nearby
    \textcolor{aoenglish}{{\bf desired branch}} predicts a 0.00 probability
    of defects.  XTREE's  plans are the delta  $\Delta$     between the  branches.}
    \label{fig:XTREE}
\end{figure}

\subsection{XTREE, 2020}

Earlier in 2020,
Krishna~\cite{krishna2017learning}  proposed XTREE, a novel defect reduction planning method that does not rely on outlier statistics. The XTREE planner uses frequent pattern mining, decision trees, and a  walk traversal algorithm.

With the pattern mining, XTREE attempts to find what code metrics usually change together by applying an association rule learner on historical data. Since metrics in Table \ref{ck} are continuous, XTREE will first discretize the values into intervals using Fayyad-Irani. Then a FP-growth algorithm \cite{han2007frequent} is used to mine frequent itemsets (in our experimentation XTREE uses $minSupport=5\% \times total\_size$ ). 

The returned maximal frequent itemsets are  used to construct a decision tree. After that, in the third part, the plans will be generated by traversing the decision tree to seek for the closest branch with highest improvement in the probability of the non-defective label. An example of the traversal procedure is illustrated in the Figure \ref{fig:XTREE}. Once the \textcolor{orange}{{\bf current branch}} is found,
  the plan will be the $\Delta$ from the current branch to a nearby  \textcolor{aoenglish}{{\bf desired branch}} with lower probability of
  defects.

\section{New Methods for  Planning Defect\newline Reduction (LIME and TimeLIME)}\label{timelime} 

\subsection{LIME}\label{lime}

\begin{figure}[!t]
\tiny
\includegraphics[width=1.0\linewidth,height=4cm]{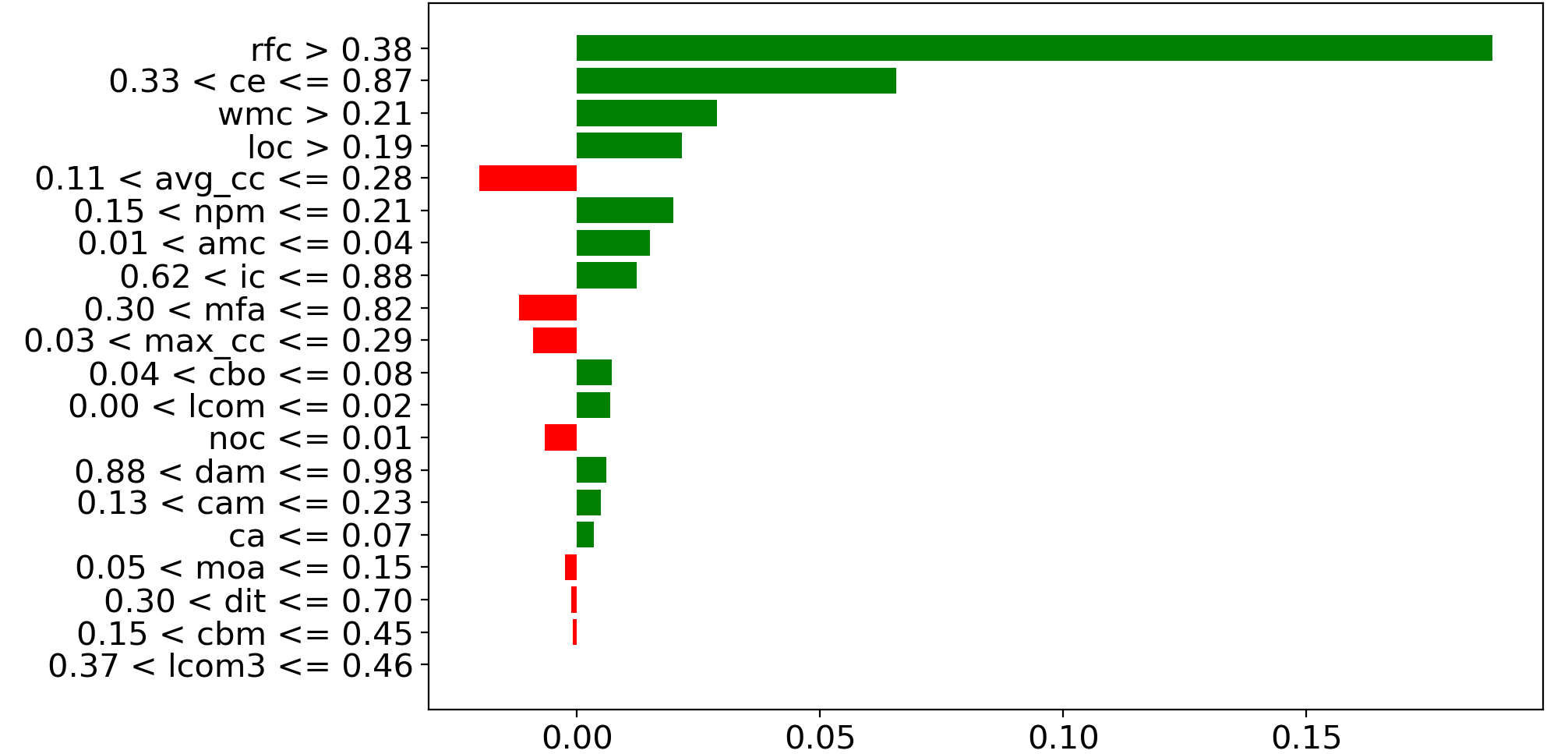}
\caption{
An example of
output generated by 
Table~\ref{inside} when applied
to the data sets of the form
of Table~\ref{ck}.
The y-axis shows the feature name and the confidence interval during which the explanation stays effective. The x-axis indicates the importance weight of each attribute. The prediction label of this instance is 1 (defective), and the weights show how each feature contributes to the prediction. A \textcolor{aoenglish} {{\bf positive}} weight means the feature encourages the classifier to predict the instance as a positive label (defective), and vice versa for the \textcolor{red} {{\bf negative}} weight. Larger weights indicate greater feature importance in terms of the prediction value based on that feature weighted by a similarity kernel.}
\label{fig:lime}
\end{figure}

One of the starting points of this research was the
realization that the LIME  algorithm, first published at KDD'16~\cite{ribeiro2016should}
could be applied to defect reduction planning. 
The internal framework of LIME is depicted in Table \ref{inside}.
In summary, 
given an instance $I$ of class $X$, LIME conducts a sensitivity analysis
in the neighborhood around $I$ to determine what could change
the class from $X$ to $Y$. 
 Using the synthetic data generated around $I$, LIME can get the classification/regression result from any black-box learner, which will then be used to fit a linear model describes the local region.  The parameters of the fitted linear model are then reported
 as a way to understand how changes in values can adjust the classification; e.g.,  see  
Figure \ref{fig:lime}.
 
\begin{figure}[b!]
\includegraphics[width=.48\textwidth]{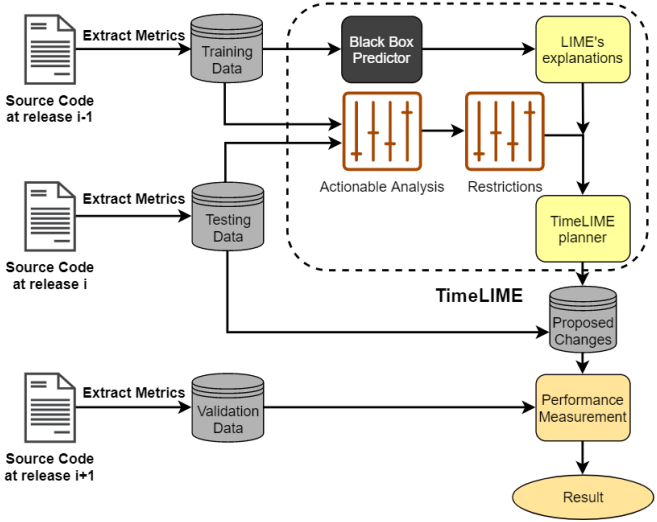}
\caption{ TimeLIME: overview of the algorithm, plus the  $K$-test evaluation rig. Note that for evaluating other benchmark algorithms, the area bounded by the dotted line will be replaced by the corresponding algorithm. For further details on TimeLIME,
see Algorithm \ref{refineflip}.} 
\label{fig:frame}
\end{figure}

This paper utilizes LIME and its capability in interpretation to generate 
defect reduction
plans.
If a black-box model can predict defects accurately, then it might be "knowledgeable" enough to provide more informative plans than a subject matter experts can provide. The key question is, therefore, how could we access the knowledge owned by a black-box model. In this paper, we imported LIME as the core component of our defect reduction algorithm as we also leverage other software domain knowledge to help LIME restrict the proposed plans in an effective fashion.

Sometimes, we are asked why we are basing
our approach on LIME and not other other tools
that explain how to change attributes in order the change
the classification of an instance.
To say the least, there are very many alternate algorithms.
 A recent survey by Mueller et al.   summarized various kinds of change-explanation generation tools.~\cite{mueller2019explanation}.
Mentioned in their study, Mueller et.al report that this literature is truly vast.   Consequently, there are many alternatives to LIME including the abductive framework of Menzies et al.~\cite{menzies02} or ANCHORS~\cite{ribeiro2018anchors} (which is another 
change-explanation algorithm generated by the same team that created LIME).

We based our work on LIME, for several reasons.
Firstly, LIME scales to large problems. Much recent work has results in methods to scale data mining to very large data sets. Since LIME is based on data mining, then LIME can use those scalability results in order to generate explanations for very large problems.

Secondly, and this is more of a low-level systems reason, alternatives to LIME such as ANCHORS assume categorical or discrete features. Our data has continuous classes which could be binarized into two discrete classes-- but only at the cost of losing the information about local gradients. Hence, at least for now, we explore LIME (and will explore
ANCHORS in future work).

Lastly, LIME is a widely-cited algorithm. At the time of this writing, LIME has received over 3,000 citations since it was published in 2016. Hence, methods used to improve LIME could also be useful for a wide range of other research tasks. This paper proposes precedence plausibility as a way to improve LIME.

\subsection{TimeLIME}\label{gpr}

TimeLIME extends LIME by restricting
the generated recommendations to the   attributes which were seen to be frequently modified within the history of a software projects. Figure~\ref{fig:frame} offers a graphical overview of this   system.

TimeLIME evolved out of comments we heard at
workshop on "Actionable Analytics" at ASE'15\cite{hihn2015data}. There, business users complained about analytic models saying that rather   applying a black-box data mining algorithm, they preferred an approach with a seemingly intuitive appeal. Since software engineers are the target audience of analytics in SE, it is crucial to ensure the proposed recommendations are valued by them.  Chen et al. say the term "actionable" can be defined as a combination of "comprehensible" and "operational"\cite{chen2018applications}. But how to assess "operational"? 

In this paper we make the following assumption about ``operational'': a proposed change to the code is plausible  if it has occurred before. That is, in this work, we claim a plan is the most operational when it has the most precedence in the history log of the project.

Using this assumption, we can generate operational analytics by:
\begin{itemize}
\item Looking at two releases of a project
and report the attributes that have changed between them; 
\item Next, when generating plans, we only used those attributes that have the most changes.
\end{itemize}
After conducting a survey on 92 controlled experiments published in 12 major software engineering journals, Kampenes et al.~\cite{Kampenes2007} argues that in SE,  size change can be measured via Hedge's $g$ value\cite{rosenthal1994parametric}:
\begin{equation}\label{g}
    g =(M_1 - M_2)/(S_{pooled} )
\end{equation}
Here,  $M_1$ and $M_2$ are the means of an attribute in two consecutive releases  and $S_{pooled}$ comes from
\ref{spool}. This expression is the pooled and weighted standard deviation ($n$ and $s$ denote the sample size and the standard deviation respectively).
\begin{equation}\label{spool}
    S_{pooled} = \sqrt{((n_1-1)s_1^{2}+(n_2-1)s_2^{2})/(n_1+n_2-2)}
\end{equation}
For the details on how Equation~\ref{spool} was applied, see \S\ref{planner}.

Furthermore, in order to ensure the precedence of plans generated by TimeLIME, we will only allow changes that have actually been "seen" in the record before.
When a plan is generated given a certain set of actionable features (we denote such set as the candidate pool), we look it up in the historical records of the project. If there exist such records where the exact changes proposed by TimeLIME had once been made by developers, we will return the plan. Otherwise, the algorithm will go back to the previous step and start generating a different plan using the same candidate pool. The plan generating process is a greedy attempt: Only if all the combinations of $M$ candidates fail, we will try to make a plan using ($M-1$) candidates.  In theory, TimeLIME might generate conflicting plans (but in our logs, this occurs very rarely). Nevertheless, to handle that situation, we recommend that if two plans are conflicting, users should adopt the plan with greatest support (i.e. most frequent in the historical log).

The reason why TimeLIME uses association rule mining to ensure the precedence of plans is that TimeLIME is designed to provide not {\bf optimal} solutions, but {\bf achievable} and {\bf maintainable} solutions. Different development teams have different capacity of reducing defects. If a planner is learning from the historical activities of developers (which contains examples of good defect reductions) to generate plans, then better plans could be generated by learning from better development teams. Thus, it is impossible to generate an optimal plan since no one can assert whether the current plan is optimal. Alternatively, it is viable to generate a maintainable plan: if the current team has been working on reducing defects in a satisfactory (may not be best) level, TimeLIME would help to maintain the defect reduction quality that has been achieved so far. 

\section{Assessing Planners:  the K-test} \label{K-test}

This paper claims that plans from TimeLIME planner  (that focus on attributes with a history of most change) outperform those generated from LIME, XTREE, Alves,
Shatnaw, and Oliveira.
To defend that claim, we need some way to assess different planning methods.

There's an expression in Latin, \emph{post hoc ergo propter hoc}, which means "after this, then because of this". This expression refers to the logical fallacy that "if event B follows event A, then event A must be the cause of event B". The assertion is obviously flawed since other events could be the true trigger of event B. This is why, in this study, we need to carefully evaluate the effectiveness of plans to see if knowledge learned from past code change records actually helps make plans on future code changes.

To address this concern, we use Krishna's $K$-test\cite{krishna2017learning}.
The $K$-test uses historical data from multiple software releases to compare the effectiveness
of  different plans $P_1,P_2,....$. The test is a kind of simulation study that assumes developers were told about a plan at some prior time. 
 Given  project  information  divided  into {\em oldest},
 {\em newer},  and
{ \em most recent}, we will   use  the
{\em oldest} data  to  determine  what  attributes  where
often changed in a project.
Then,  using
the {\em newer} data, we will build plans using   LIME, TimeLIME, XTREE, Alves,
Shatnaw, and Oliveira. Finally, we will divide the changes
between the {\em newer} and the 
{\em most recent} into the changes that
 {\em overlap} with the plans, and those that do not.

More precisely, we use consecutive releases $x,y,z$ of some software system.
These releases are required to contain named regions of code $C_1,C_2,$ etc. that can be found in releases
$x,y,z$. For example, $C_i$ could be an object-oriented  class or a function or a file that is found in all releases.
The $K$-test then assumes that there exists a quality measure $Q$ that reports the value of the regions of named code in different releases. In this study, we will use NDPV (\emph{Number of Defects in Previous Version}) as the quality measure, which is described later in \S\ref{Performance Criteria}. 
Some method is then applied that uses $Q$ to reflect on the releases $x,y$
in order to infer a plan $P_i$ for improving release $z$\footnote{Note the connection here to temporal validation in machine learning~\cite{Witten:2011}. In the $K$-test, no knowledge of the final release $z$ is used to generate the plans.}.

Given the above, the $K$-test collects following quantities to address our claims made in introduction:
\begin{itemize}[leftmargin=16pt,topsep=4pt]
\item 
\textbf{RQ1: Smaller}: To measure the succinctness of plans, we collect the number of changes within the plan proposed for each code $C_i$ in release $y$. 
\item 
\textbf{RQ2: Ready to apply}: To measure how likely a plan can be realized by developers, we compute $J_{y,z}= \Delta_{y,z} \cap P_i$: the overlap between the proposed plan and the code changes.
\item
\textbf{RQ3: Better}: To measure which planner is better at reducing defects, we collect $Q_z-Q_y$: i.e.,  the change in the number of bugs of the named code $C_i$ between releases $y,z$. Then, we weight the change $Q_z-Q_y$ by $J_{y,z}$. The intuition is that the planner cannot get credit in a bug-reducing code file if its plan shares little or none similarity with the actual actions done by developers.
\end{itemize}
The $K$-test defines \emph{better} plans as follows:
\begin{quote}{\bf DEFINITION:} {\em
 Plan $P_i$  is ``better''
 that     plan $P_j$   if,
  in release $z$, $P_i$ is
  associated with most
  quality improvements.}
 \end{quote}
 That is,  
  increasing the size of the overlap of the proposed plan   is associated with increasing quality in release $z$; i.e., 
\[(Q_z-Q_y) \propto |J_{y,z}| \]
 For our purposes, the $K$-test procedure in this paper consists of three steps:
\begin{itemize}
    \item Train a defect reduction planner on version $x$.
    
    \item  Use trained planner to generate plans with the aim of fixing bugs reported in version $y$.  In this step, classical LIME planner and TimeLIME planner will utilize the explanations from the explainer and TimeLIME, in addition, will use the historical data analysis to generate plans.
    
    \item On the same set of files that are reported buggy in version $y$, we measure $J_{y,z}$, the overlap score of each plan and the changes in the version $z$, using the Jaccard similarity function. We also record $Q_z-Q_y$, the change in the number of bugs between the version $y$ and version $z$.
\end{itemize}
For each instance, we compare the extent of overlap between the recommended plan $P_i$ generated by the planner   and the actual developer action in the next release as $\Delta_{y,z}$ using the Jaccard similarity coefficient. 
\begin{equation}\label{similar}
    J_{y,z}(P_i,\Delta_{y,z}) = (P_i \cap \Delta_{y,z})/(P_i \cup \Delta_{y,z})
\end{equation}
Then we convert the coefficient into percentage as our overlap score. As an example shown in Table \ref{tab:overlap}, the overlap score is \[2/4 \times 100\% = 50\%\]


\begin{table}[!h]
\centering
\begin{tabular}{|r|c|c|c|c|} 
\hline
 & AMC      & LOC        & LCOM        & CBO  \\ \hline
Current release y     & 0.2   & 0.1     & 0.9  & 0.5   \\ 
\hline
$P_i$ for release z       & {\cellcolor[rgb]{0.753,0.753,0.753}}no change & {[}0, 0.1] & {\cellcolor[rgb]{0.753,0.753,0.753}}{[}0, 0.9] & no change  \\ 
\hline
Next release z & {\cellcolor[rgb]{0.753,0.753,0.753}}0.2          & 0.3        & {\cellcolor[rgb]{0.753,0.753,0.753}}0.3          & 0.2           \\
\hline
Match? & y & n & y & n\\\hline
Map to Table \ref{tab:cm} & TN & FP & TP & FN\\ \hline

\end{tabular}
\caption{A contrived example: computing  similarity score using the Jaccard   function from Equation (\ref{similar}). Plans that match the developer actions are marked gray.}
\label{tab:overlap}
\end{table}

\begin{table*}[!b]
\centering
\begin{tabular}{|l|l|l|l|l|l|l|l|l|} 
\hline
 & Training & Testing & Validation& No. of & No. of  &No. of bugs & No. of bugs & No. of\\
Dataset  & (oldest) & (newer) & (most recent) & files & matched files & in testing set & in validation set &  bugs reduced \\ 
\hline
Jedit    & 4.0 & 4.1 & 4.2   & 367  &78  & 216 & 74 & 142 \\
Camel1    & 1.0  &  1.2  &  1.4   & 872 & 210 & 508 & 247 & 261  \\
Camel2    & 1.2  &  1.4  &  1.6   & 965  & 144 & 334 & 316 & 18 \\
log4j    & 1.0   & 1.1    & 1.2  & 205   & 35 & 83 & 120 & -37\\
Xalan    & 2.5 &  2.6 &  2.7   & 885     & 385 & 529 & 381 & 148 \\
Ant      & 1.5 & 1.6 &  1.7   & 745     & 91 & 183 & 163 & 20\\
Velocity & 1.4 &  1.5 &  1.6   & 229    & 138 & 321 & 144 & 177 \\
Poi      & 1.5 &  2.5 &  3.0   & 442    & 247 & 495 & 366 & 129 \\
Synapse  & 1.0 &  1.1 &  1.2   & 256    & 58  & 97 & 65 & 32 \\
\hline
\end{tabular}
\caption{Defect datasets used in this paper. Each columns represent a different dataset.
The last row shows the total number of bugs reduced among the same files between the testing release and the validation release. Note that a negative value in this column indicates that the validation release contains more bugs than the previous one.}
\label{tab:dataset}
\end{table*} 
Formally speaking, the $K$-test is {\em not}
a deterministic  statement that some plan will necessarily
improve quality in some future release of a project.  Such deterministic causality is a precisely defined concept with the property that a single counterexample can refute the causal claim~\cite{AAAI_1990}. The $K$-test does \underline{not} make such statements.

Instead, the $K$-test is a statement of historical  observation. Plans that are ``better'' (as defined above)
are those which, in the historical log, have been
associated with increased values on some quality measure. Hence, they have some likelihood (but no certainty) that they will do so for future projects.

\section{Experimental methods}\label{experiment}
The experiment reports the performance of TimeLIME and other state-of-the-art works by comparing the quality of plans recommended by each method.

Firstly, we use an over-sampling tool called SMOTE\cite{chawla2002smote} to transform the imbalanced datasets in which defective instances may only take a small ratio of the population. This was needed since, in many of the prior papers that explored our data,  researchers warn that small target classes made it harder to build predictors~\cite{amrit18}.

Secondly, as discussed above, we train the predictor $P$ and explainer $E$ on data of version $x$. Then in version $y$ we use the explainer to generate explanations {\em only} on those data that are reported as buggy. We also use the predictor $P$ to determine whether we should provide recommendation plans to the instance. 

Then we measure the overlap score of our recommended plan and the actual change on the same file in version $z$. To do this, only select instances that are defective and whose file name has appeared in all releases of data to be instances in need of plans. 

The above steps are applied for each benchmark method as well as the TimeLIME planner proposed by this paper. The visualization of the experimental rig is shown in Figure \ref{fig:frame}. In the classical LIME planner, we use the simple strategy which is to change as many features as it can in order to reduce the number of bugs. On the other hand, for TimeLIME, we first input historical data from the older release to compute the variance of each feature. Then we selected the top-$M$ features with the largest variance as \emph{precedented} features, meaning any recommendation on other features will be rebutted. After getting recommended plans from both planners, we assess the performance of two planners using the overlap score as described in \S\ref{Performance Criteria}. 

Note that the parameter $M$ can be user-specified and the features may vary with respect to different projects and the releases used as historical data. Here we set the default value of $M$ to be 5, which means only $25\%$ of all twenty features can be mutated. Our results from experiments suggest that $M=5$ is a useful default setting. Future work shall explore and compare other values of $M$.

\subsection{Data}
To empirically evaluate classical LIME vs TimeLIME, we use the standard datasets and measures widely used in defect prediction. In this paper, we selected 8 datasets from the publicly available SEACRAFT project\cite{jureczko2010towards} collected by Jureczko et al. for open-source JAVA systems (\url{http://tiny.cc/defects}). These datasets keep the logs of past defects as shown in Table \ref{tab:dataset} and summarize software components using the CK code metrics as shown in Table \ref{ck}. Note that all the metrics are numerical and can be automatically collected for different systems\cite{nagappan2005static}. The definition and nature of each attribute in the metrics is elaborated by prior researchers Jureczko and Madeyski \cite{jureczko2011significance,madeyski2015process}. Another reason this paper selects these 8 datasets is that they all contain at least 3 consecutive releases, which is required by the evaluation measure described in \S\ref{K-test}. Since Camel dataset contains 4 consecutive releases, the experiment has 9 trials in total.

\subsection{ Learner}
While other benchmark algorithms don't need the predictive learn within their model, LIME does require the user to pass in the customized learner, which can be used to generate explanations. Since the goal of this paper is to examine the performance of the defect reduction tools rather than the predictive model, this paper takes one classifier to apply the explanation algorithm on.

Our choice of classifier is guided by the  Ghotra et al. \cite{ghotra2015revisiting} study that explored  30 classification techniques for   defect prediction. They found
that all the classifiers they explored fell into four groups
and that  Random Forest classifiers were to be found in their top-ranked group.

A Random Forest classifier is an ensemble learner that fits a number of decision tree classifiers on different sub-samples of the dataset and generates predictions via average voting from all the classifiers\cite{ho1995random}. It is impossible to visualize a fitted Random Forest classifier as a finite set of rules and conditions due to the voting process. Therefore, Random Forest classifier is considered a non-interpretable model. Hence, it is a suitable choice for this study.

\subsection{Planners} \label{planner}
This section discusses the internals of our planners,
including a RandomWalk planner (which we use  to compare our results
against a baseline random guesser).
 
Using LIME, we generate plans to reduce
    classifications. We use the default parameter setting of LIME, which is 5000 samples around the instance neighborhood, and the entropy-based discretizer. The explanation object return by a LIME explainer is a tuple in which each element contains the feature name and the corresponding feature importance. It also provides a discretized interval indicating the range of values during which the feature will maintain the same effect to the prediction result. As described in Algorithm \ref{simpleflip}, the simple planner based on the classical LIME will recommend changes on all features that contribute to the defective prediction. 
Algorithm \ref{refineflip} shows the TimeLIME planner, 
which utilizes Algorithm \ref{findsupport} to ensure that the proposed plan must be precedented in he historical records.
Each planner
uses feature ranges generated by flipping the discretized interval relative to the feature value range $[0,1]$.

Also, just to compare, we   use a planner named RandomWalk as a ``straw-man'' baseline algorithm.  This planner, as shown in Algorithm \ref{randomwalk}, assigns random recommendations to each variable stochastically.

One final note: to make the comparisons fair, in our experiment setting, we set the number of changed features as the same as the TimeLIME planner for comparison purpose.

\begin{algorithm}[!t]
\caption{ClassicalPlanner (LIME)\label{simpleflip}}
\small
\KwData{explanation $e$ // the explanation from Table \ref{inside}}
\KwResult{A tuple consisting of intervals of values $v'$}
\Begin{
$w, v$ $\gets$ $e$ // split weights $w$ and value intervals $v$ from $e$\\
$i \gets0$\\
\While{$i \leq$ sizeof $(w)$ }{
  \eIf{$w[i] \geq$ 0}{
   $v'[i] \gets flip (v[i])$\\
   }
   {
   $v'[i] \gets v[i]$  // do not propose a change on this feature
  }
  $i \gets i+1$
  
}
return $v'$
}
\end{algorithm}


\begin{algorithm}[t] 
\caption{TimeLIME Planner\label{refineflip}}
\small
\KwData{explanation $e$ from Table \ref{inside}, precedence parameter $M$, previous release $x$, current release $y$}
\KwResult{A tuple consisting of intervals of values $v'$}
\Begin{
$w, v \gets e$ // split weights $w$ and value intervals $v$ from $e$\\
$M \gets$ 5 // the default parameter $M$ is 5 meaning at most 5 features can be changed in the resulting plan\\
$g \gets$ hedge($x,y$) // defined in  \S\ref{gpr} \\
precedented $\gets sorted (g) [0:M]$\\
$i \gets0$\\
$pool \gets v$\\
\While{$i \leq$ sizeof $(w)$}{
  \eIf{$w[i] \geq$ 0 and $i \in$ precedented}{
   $pool[i] \gets flip(v[i])$\\
   }
   {
   $continue$ // do not propose a change
  }
  $i \gets i+1$
}
$v' \gets findSupport(pool,precedented,x,y)$ \\// as described in Algorithm \ref{findsupport}\\
return $v'$
}
\end{algorithm}

\begin{algorithm}[t] 
\caption{{\em findSupport}
\label{findsupport}}
\small
\KwData{candidates changes on each single feature $pool$, precedented features $precedented$, previous release $x$, current release $y$}
\KwResult{The proposed plan for the instance}
\Begin{
$max\gets 0$ // initialize the max support \\
$itemsets\gets (x,y)$ // get records of actual changes \\
$M \gets5$ // number of changes allowed\\
\While{$M>0$}{
  $plans \gets generatePlans(pool, M, precedented)$\\
  $plan, max \gets findMax(plans, itemsets)$ // return the plan with max support in the historical records\\
  \eIf{$max >$ 0}{
   $break$\\
   }
   {
   $M = M-1$ // try to propose fewer changes\\
  }
}
return $plan$
}
\end{algorithm}

\begin{algorithm}[t] 
\caption{RandomWalk\label{randomwalk}}
\small
\KwData{ standardized code instance to be explained $c$,number of features to be mutated $n$}
\KwResult{A tuple consisting of intervals of values $v$}
 \Begin{
$pool \gets random.sample(20,n)$ // randomly choose n out to 20 features\\ 
$i\gets0$\\
\While{$i \leq$ sizeof $(c)$ }{
  \eIf{$p in$ pool}{ 
  $(a,b) \gets$ $sorted$(rand(1),rand(1)) // generate a random interval within the range [0, 1].\\
   $v[i] \gets (a,b)$ // apply the random interval.\\ 
   }
   {
   $v[i] \gets c[i]$\\
  }
  $i \gets i+1$
}
return $v$
}
\end{algorithm}

\subsection{Performance Criteria} \label{Performance Criteria}
The two performance criteria in this experiment, as described in the \S\ref{K-test}, are the overlap score of individual plans and the number of bugs reduced/added in the next release of the project. The function used for computing the overlap score is the Jaccard similarity function in Eq. \ref{similar}, and the other criterion is measured by the metric NDPV (\emph{Number of Defects in Previous Version}), which returns the number of bugs fixed (or added) in a given file during the development of the previous release. The nature of NDPV and similar metrics have been evaluated by plentiful researchers\cite{jureczko2010using,couto2014predicting,shihab2010understanding,khoshgoftaar1998using}. 

\begin{figure}[!t]
\centering
\includegraphics[width=.48\textwidth]{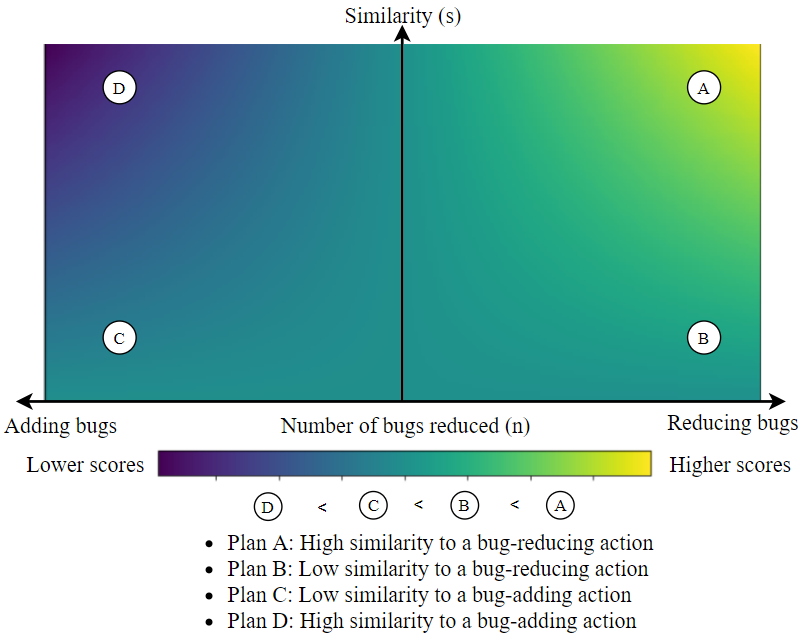}
\caption{
Visualized tendency of the Eq. \ref{ws}. The x-axis shows the NDPV $n_i$ and the y-axis shows the similarity score $s_i$. \BLACK} 
\label{fig:example}
\end{figure}

To further evaluate the second criterion, we chose to use a weighted sum function to compute the net gain of each planner. The weighted sum function in Eq. (\ref{ws}) weights the NDPV by the overlap score of the plan. 
\begin{equation}\label{ws}
    S= \sum {s_i * n_i}
\end{equation}
In the experiment, each plan $p_i$ from the all $N$ plans returns an overlap score $s_i$ and a
NDPV number $n_i$ (positive number indicates bugs reduced, negative number indicates bugs added).
Then we multiply the NDPV $n_i$ by $s_i$ to compute the weighted improvement score $S$. Note that the {\em larger} $s_i$ indicate the {\em greater} overlap. 
Fig. \ref{fig:example} shows the tendency of the weighted improvement score $S$ with respect to $s_i$ and $n_i$. 
Generally speaking, we will reward plans who are similar to a bug-reducing change and penalize those plans who are similar to a bug-introducing change. In the case where the plan is very dissimilar to a change (whether it is bug-reducing or not), we assign a trivial score to the plan since it shares little overlap with the actual change, which makes it impossible for us to simulate the potential consequence of applying such plan.   
  
Additionally, given that the total number of bugs varies from each project as shown in Table \ref{tab:dataset}, a project with more bugs reduced in the validation dataset will expect the planner to score more than the planner whose validation dataset has fewer bugs reduced so that their performance can be considered proportionally similar. For example, project A has $NDPV = 100$ in release $y$ and another project B has $NDPV = 10$ in its next release y. Assume one would like to observe similar performance of a planner on these 2 projects, 
it won't make any sense if the planner gains the same score in both projects.  
From this perspective, we scale the final score $S$ in Eq. \ref{ws} by the sum of NDPV within the project to get the scaled score $S_{scaled}$.
\begin{equation}\label{scaled}
    S_{scaled} =  \frac{\sum_{i}^{N} {s_i * n_i}}{  \sum_{i}^{N} {n_i} }
\end{equation}

\begin{table*}[!t]
\begin{adjustbox}{max width=\textwidth}
\begin{tabular}{|l|l|l|l|l|l|l|l|l|l|l|l|}
\hline
 &
  Jedit &
  Camel1 &
  \multicolumn{2}{c|}{Camel2} &
  Log4j &
  \multicolumn{2}{c|}{Xalan} &
  Ant &
  Velocity &
  Poi &
  Synapse \\ \hline
TimeLIME Plans &
  - lcom3 &
  - lcom3 &
  + lcom3 &
  - moa &
  - max\_cc &
  - cbo &
  - loc &
  - rfc &
  - cbm &
  + rfc &
  + avg\_cc \\
 &
  - moa &
  + dam &
  + moa &
  - ce &
  - avg\_cc &
  + loc &
  + cam &
  - ce &
  - mfa &
  + amc &
  - cbm \\
 &
  - avg\_cc &
  + cam &
  - ce &
  - lcom3 &
  -wmc &
  + amc &
  + cbo &
  - npm &
  - amc &
  - dam &
  - mfa \\
 &
  - max\_cc &
  - ic &
  + rfc &
  - rfc &
  - npm &
  - cam &
  + ce &
  -wmc &
   &
  + loc &
  - cam \\
 &
  + cam &
   &
   &
   &
   &
   &
   &
   &
   &
   &
   \\ \hline
Refactoring Methods &
  \circled{2}, \circled{3}, \circled{15} &
  \circled{3}, \circled{15} &
  \circled{4},\circled{16} &
  \circled{3}, \circled{5}, \circled{15} &
  \circled{3}, \circled{5}, \circled{8}, \circled{9}&
  \circled{4}, \circled{7} &
  \circled{3} &
  \circled{1}, \circled{5}, \circled{13} &
  \circled{1}, \circled{7}, \circled{8}, \circled{9}&
  \circled{4}, \circled{12}, \circled{16} &
  \circled{4}, \circled{13} \\
   &
   &
   &
   &
   &  \circled{11}, \circled{13}, \circled{15}
   &
   &
   &
   &  \circled{11}, \circled{13}, \circled{15}
   &
   & \\
  \hline
\end{tabular}
\end{adjustbox}
\caption{ Using Table \ref{tab:refactoring}, developers can map TimeLIME's plans onto some simple refactoring methods to achieve the desired changes in code metrics. The "+" and "-" indicate increase and decrease respectively. Note that while various plans are provided within each project, in this table we only show the most frequent plan(s).  
}
\label{tab:mapping}
\end{table*}

\section{Results}\label{result} 

\subsection{RQ1: Does TimeLIME provide succinct plans?}

\begin{figure}[!t]
    \includegraphics[width=1.02\linewidth]{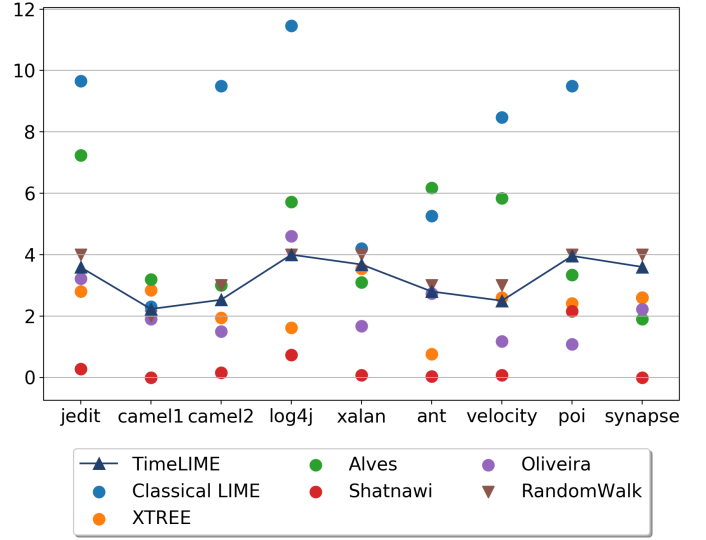}
    \caption{{\bf RQ1 results}: The mean size of TimeLIME's plans 
    (across all instances in release $z$) is often much smaller than LIME.  Y-axis shows the number of features changed by recommended plans. }
    \label{fig:size}
\end{figure}

Figure \ref{fig:size} reports the 
mean size of plans across all instances in release $z$.
We note that:
\bi
\item
RandomWalk method's plans are so large since this planner
does not use information from the domain to constraint its results.
\item
TimeLIME generates much smaller plans compared to many other planners including  classical LIME.
\item
The only planner that consistently produces smaller plans in the Shatnawi method but,
as seen in the {\b RQ3} results (below), the Shatnawi obtains performance that is far worse than TimeLIME.
\ei
Note that since TimeLIME in the experiment restricts plans to the top 5 features with highest Hedge's $g$ scores, the size of an TimeLIME plan will never be more than 5. However, as shown in the figure, the average size of TimeLIME plans is always smaller than 5. This implies that the original code refactoring plans, proposed by the classical LIME planner, do contain unprecedented changes which then get rejected by the TimeLIME planner.   
In summary, for {\bf RQ1}, we say:
\begin{blockquote}
\textbf{Answer 1}: 
     Several planners, including LIME, generate plans that are far larger than those found by TimeLIME. And the only planner that always generates smaller plans has much worse performance.
\end{blockquote}


\subsection{RQ2: Could developers apply the changes proposed by TimeLIME?}\label{rq2better}

We answer this question in two ways.
  Firstly, we assess  ``can developers map our plans onto known refactoring actions?'' (and for our definition of ``refactoring actions'', please see Table~\ref{tab:refactoring}).  
 Table~\ref{tab:mapping} shows those mappings.
While
 things get somewhat  complicated in two cases (Log4j and Velocity) it is encouraging to note that in $\frac{9}{11}$ cases,
the number of refactoring actions is {\em less than}
the number of changes recommended in TimeLIME's plans.
Concrete examples of how to apply these mapped plans can be found on our homepage. \footnote{https://github.com/ai-se/TimeLIME/blob/master/README.md\#examples}. 
 
  For a second way to answer this question, we use our historical data.
 We posit the scenario
that developers were told of our plans in the {\em current} release,
and then we check the {\em later} release to see if the plans we proposed were actually recommended.
As shown in the rest of this section,
we say our plans are {\em feasible}  since there is evidence indicating developers could actually apply those changes.

Table \ref{tbl:med_overlap} and Table \ref{tbl:iqr_overlap} comments on how often developers are willing to perform the plans suggested by different planners. Both tables are generated using the $K$-test procedure described above. Each cell in Table \ref{tbl:med_overlap} shows the median value of the
$J_{y,z}$ overlap score measured from Eq. \ref{similar} in \S\ref{K-test} for all instances within the projects. In addition, in order to explore the robustness of our approach, we added 2 variants of TimeLIME planners, each embedded with a different predictor. While the original TimeLIME planner uses a random forest classifier internally, the 2 variants use multi-layer perceptron (MLP) and support vector machine (SVM) respectively.

Table \ref{tbl:iqr_overlap} shows the interquartile range (IQR) of all overlap scores quantile among all plans generated by planners. With similar median scores, a smaller IQR means the planner is more stable and robust. It is noteworthy that the Random Planner always obtains very small IQRs in all project. This is because plans generated by Algorithm \ref{randomwalk} are equivalently bad as indicated from the median scores. On contrary, TimeLIME has similarly small IQRs while maintaining the highest median scores in all project, which means it prevails other planners in terms of providing plans that better resemble developers' choices. In summary:

\begin{itemize}[leftmargin=16pt,topsep=4pt]
\item Unsurprisingly, Random Planner has the lowest similarity scores in all projects.
\item
The 4 prior works (XTREE, Alves, Shatnawi, and Oliveira) are equivalently good. 
\item
Different projects have very different baselines for the similarity evaluation. For example, within Xalan, every planner except Random obtains a relatively high scores whereas they perform equally poorly in the Ant project.
\item
All TimeLIME planners have obtained the highest score in every project with a relatively low IQR scores. This means the performance of TimeLIME, regardless of the type of the embedded classifier, is good and robust in terms of similarity to actual actions.
\item
The classical LIME planner has a volatile performance: It is either performs   best or  worst. In other words, compared to TimeLIME, we cannot recommend that procedure for practical purposes. 
\end{itemize}
In summary, we answer {\bf RQ2} as follows:
 \begin{blockquote}
\noindent
\textbf{Answer 2}: We find a large   overlap  between  TimeLIME’s  recommendations  and  the possible actions (Table~\ref{tab:mapping}) and  observed actions (Table~\ref{tbl:med_overlap}, Table~\ref{tbl:iqr_overlap}) of  developers.
\end{blockquote}

Hence we say developers would be able to apply TimeLIME's recommendations.

\begin{table}[!h]
\centering
    
\begin{tabular}{|r|r|r|r|r|r|r|r|r|r|}
\hline
{\color[HTML]{000000} } & {\color[HTML]{000000} \rotatebox{90}{Random} } & {\color[HTML]{000000} \rotatebox{90}{Oliveira}} & {\color[HTML]{000000} \rotatebox{90}{Shatnawi}} & {\color[HTML]{000000} \rotatebox{90}{Alves}} & {\color[HTML]{000000} \rotatebox{90}{XTREE}} &{\color[HTML]{000000} \rotatebox{90}{LIME}} & {\color[HTML]{000000} \rotatebox{90}{TimeLIME}} & {\color[HTML]{000000} \rotatebox{90}{TimeLIME\_MLP}} & {\color[HTML]{000000}  \rotatebox{90}{TimeLIME\_SVM}} \\
\hline
{\color[HTML]{000000} Jedit} & \cellcolor[HTML]{FFFFFF}{\color[HTML]{000000} 30 } & \cellcolor[HTML]{FFFFFF}{\color[HTML]{000000} 35 } & \cellcolor[HTML]{FFFFFF}{\color[HTML]{000000} 35 } & \cellcolor[HTML]{FFFFFF}{\color[HTML]{000000} 30 } &
\cellcolor[HTML]{F4F4F4}{\color[HTML]{000000} 40 } &
\cellcolor[HTML]{FFFFFF}{\color[HTML]{000000} 35 } & \cellcolor[HTML]{A9A9A9}{\color[HTML]{000000} 90 } & \cellcolor[HTML]{A9A9A9}{\color[HTML]{000000} 85 } & \cellcolor[HTML]{A9A9A9}{\color[HTML]{000000} 90 } \\
{\color[HTML]{000000} Camel1} & \cellcolor[HTML]{E6E6E6}{\color[HTML]{000000} 55 } & \cellcolor[HTML]{E6E6E6}{\color[HTML]{000000} 63 } & \cellcolor[HTML]{E6E6E6}{\color[HTML]{000000} 65 } & \cellcolor[HTML]{E6E6E6}{\color[HTML]{000000} 60 } &
\cellcolor[HTML]{E6E6E6}{\color[HTML]{000000} 67.5 } &   
\cellcolor[HTML]{919191}{\color[HTML]{000000} 85 } & \cellcolor[HTML]{919191}{\color[HTML]{000000} 95 } & \cellcolor[HTML]{919191}{\color[HTML]{000000} 95 } & \cellcolor[HTML]{919191}{\color[HTML]{000000} 95 } \\
{\color[HTML]{000000} Camel2} & \cellcolor[HTML]{F4F4F4}{\color[HTML]{000000} 45 } & \cellcolor[HTML]{E6E6E6}{\color[HTML]{000000} 55 } & \cellcolor[HTML]{E6E6E6}{\color[HTML]{000000} 55 } & \cellcolor[HTML]{E6E6E6}{\color[HTML]{000000} 50 } &
\cellcolor[HTML]{E6E6E6}{\color[HTML]{000000} 50 } &   
\cellcolor[HTML]{F4F4F4}{\color[HTML]{000000} 40 } & \cellcolor[HTML]{919191}{\color[HTML]{000000} 95 } & \cellcolor[HTML]{919191}{\color[HTML]{000000} 95 } & \cellcolor[HTML]{919191}{\color[HTML]{000000} 95 } \\
{\color[HTML]{000000} Log4j} & \cellcolor[HTML]{F4F4F4}{\color[HTML]{000000} 45 } & \cellcolor[HTML]{E6E6E6}{\color[HTML]{000000} 50 } & \cellcolor[HTML]{F4F4F4}{\color[HTML]{000000} 45 } & \cellcolor[HTML]{F4F4F4}{\color[HTML]{000000} 40 } &
\cellcolor[HTML]{E6E6E6}{\color[HTML]{000000} 50 } &
\cellcolor[HTML]{FFFFFF}{\color[HTML]{000000} 35 } & \cellcolor[HTML]{A9A9A9}{\color[HTML]{000000} 75 } & \cellcolor[HTML]{919191}{\color[HTML]{000000} 80 } & \cellcolor[HTML]{A9A9A9}{\color[HTML]{000000} 75 } \\

{\color[HTML]{000000} Xalan} & 
\cellcolor[HTML]{A9A9A9}{\color[HTML]{000000} 70 } & \cellcolor[HTML]{919191}{\color[HTML]{000000} 85 } & \cellcolor[HTML]{919191}{\color[HTML]{000000} 90 } & \cellcolor[HTML]{919191}{\color[HTML]{000000} 80 } & 
\cellcolor[HTML]{A9A9A9}{\color[HTML]{000000} 75 } &
\cellcolor[HTML]{919191}{\color[HTML]{000000} 85 } & \cellcolor[HTML]{919191}{\color[HTML]{000000} 100 } & \cellcolor[HTML]{919191}{\color[HTML]{000000} 100 } & \cellcolor[HTML]{919191}{\color[HTML]{000000} 100 } \\

{\color[HTML]{000000} Ant} & 
\cellcolor[HTML]{FFFFFF}{\color[HTML]{000000} 30 } & \cellcolor[HTML]{FFFFFF}{\color[HTML]{000000} 35 } & \cellcolor[HTML]{FFFFFF}{\color[HTML]{000000} 35 } & \cellcolor[HTML]{FFFFFF}{\color[HTML]{000000} 35 } &
\cellcolor[HTML]{FFFFFF}{\color[HTML]{000000} 35 } &
\cellcolor[HTML]{A9A9A9}{\color[HTML]{000000} 70 } & \cellcolor[HTML]{919191}{\color[HTML]{000000} 85 } & \cellcolor[HTML]{919191}{\color[HTML]{000000} 85 } & \cellcolor[HTML]{919191}{\color[HTML]{000000} 85 } \\

{\color[HTML]{000000} Velocity} & \cellcolor[HTML]{A9A9A9}{\color[HTML]{000000} 60 } & \cellcolor[HTML]{A9A9A9}{\color[HTML]{000000} 75 } & \cellcolor[HTML]{A9A9A9}{\color[HTML]{000000} 75 } & \cellcolor[HTML]{E6E6E6}{\color[HTML]{000000} 55 } &
\cellcolor[HTML]{A9A9A9}{\color[HTML]{000000} 65 } &
\cellcolor[HTML]{E6E6E6}{\color[HTML]{000000} 50 } & \cellcolor[HTML]{919191}{\color[HTML]{000000} 100} & \cellcolor[HTML]{919191}{\color[HTML]{000000} 95 } & \cellcolor[HTML]{919191}{\color[HTML]{000000} 100 } \\

{\color[HTML]{000000} Poi} & 
\cellcolor[HTML]{FFFFFF}{\color[HTML]{000000} 35 } & \cellcolor[HTML]{F4F4F4}{\color[HTML]{000000} 40 } & \cellcolor[HTML]{FFFFFF}{\color[HTML]{000000} 35 } & \cellcolor[HTML]{F4F4F4}{\color[HTML]{000000} 40 } &
\cellcolor[HTML]{F4F4F4}{\color[HTML]{000000} 45 } &
\cellcolor[HTML]{F4F4F4}{\color[HTML]{000000} 45 } & \cellcolor[HTML]{A9A9A9}{\color[HTML]{000000} 75 } & \cellcolor[HTML]{A9A9A9}{\color[HTML]{000000} 75 } & \cellcolor[HTML]{A9A9A9}{\color[HTML]{000000} 75 } \\

{\color[HTML]{000000} Synapse} & \cellcolor[HTML]{FFFFFF}{\color[HTML]{000000} 35 } & \cellcolor[HTML]{F4F4F4}{\color[HTML]{000000} 43 } & \cellcolor[HTML]{F4F4F4}{\color[HTML]{000000} 45 } & \cellcolor[HTML]{F4F4F4}{\color[HTML]{000000} 40 } &
\cellcolor[HTML]{E6E6E6}{\color[HTML]{000000} 50 } &
\cellcolor[HTML]{A9A9A9}{\color[HTML]{000000} 73 } & \cellcolor[HTML]{A9A9A9}{\color[HTML]{000000} 75 } & \cellcolor[HTML]{A9A9A9}{\color[HTML]{000000} 75 } & \cellcolor[HTML]{A9A9A9}{\color[HTML]{000000} 75 }\\
\hline

\end{tabular}

\caption{
{\bf RQ2 results}:
 Median overlap scores in percentage: larger scores are better, marked in darker color.
 \\}
\label{tbl:med_overlap}
\end{table}

\begin{table}[!h]
\centering
   
\begin{tabular}{|r|r|r|r|r|r|r|r|r|r|}
\hline
{\color[HTML]{000000} } & {\color[HTML]{000000} \rotatebox{90}{Random} } & {\color[HTML]{000000} \rotatebox{90}{Oliveira}} & {\color[HTML]{000000} \rotatebox{90}{Shatnawi}} & {\color[HTML]{000000} \rotatebox{90}{Alves}} & {\color[HTML]{000000} \rotatebox{90}{XTREE}} &{\color[HTML]{000000} \rotatebox{90}{LIME}} & {\color[HTML]{000000} \rotatebox{90}{TimeLIME}} & {\color[HTML]{000000} \rotatebox{90}{TimeLIME\_MLP}} & {\color[HTML]{000000}  \rotatebox{90}{TimeLIME\_SVM}}\\
\hline
{\color[HTML]{000000} Jedit} & \cellcolor[HTML]{A9A9A9}{\color[HTML]{000000} 30 } & \cellcolor[HTML]{A9A9A9}{\color[HTML]{000000} 20 } & \cellcolor[HTML]{A9A9A9}{\color[HTML]{000000} 20 } & \cellcolor[HTML]{919191}{\color[HTML]{000000} 15 } &
\cellcolor[HTML]{919191}{\color[HTML]{000000} 15 } &
\cellcolor[HTML]{F4F4F4}{\color[HTML]{000000} 50 } & \cellcolor[HTML]{919191}{\color[HTML]{000000} 15 } & \cellcolor[HTML]{919191}{\color[HTML]{000000} 19 } & \cellcolor[HTML]{919191}{\color[HTML]{000000} 15 } \\
{\color[HTML]{000000} Camel1} & \cellcolor[HTML]{F4F4F4}{\color[HTML]{000000} 45 } & \cellcolor[HTML]{F4F4F4}{\color[HTML]{000000} 50 } & \cellcolor[HTML]{F4F4F4}{\color[HTML]{000000} 55 } & \cellcolor[HTML]{F4F4F4}{\color[HTML]{000000} 45 } &
\cellcolor[HTML]{F4F4F4}{\color[HTML]{000000} 40 } &   
\cellcolor[HTML]{A9A9A9}{\color[HTML]{000000} 20 } & \cellcolor[HTML]{919191}{\color[HTML]{000000} 15 } &   
\cellcolor[HTML]{A9A9A9}{\color[HTML]{000000} 20 } & \cellcolor[HTML]{919191}{\color[HTML]{000000} 15 } \\
{\color[HTML]{000000} Camel2} & \cellcolor[HTML]{F4F4F4}{\color[HTML]{000000} 40 } & \cellcolor[HTML]{F4F4F4}{\color[HTML]{000000} 50 } & \cellcolor[HTML]{F4F4F4}{\color[HTML]{000000} 46 } & \cellcolor[HTML]{F4F4F4}{\color[HTML]{000000} 45 } &
\cellcolor[HTML]{F4F4F4}{\color[HTML]{000000} 50 } &   
\cellcolor[HTML]{A9A9A9}{\color[HTML]{000000} 26 } & \cellcolor[HTML]{919191}{\color[HTML]{000000} 10 } &   
\cellcolor[HTML]{A9A9A9}{\color[HTML]{000000} 25 } & \cellcolor[HTML]{919191}{\color[HTML]{000000} 10 } \\
{\color[HTML]{000000} Log4j} & \cellcolor[HTML]{A9A9A9}{\color[HTML]{000000} 28 } & \cellcolor[HTML]{F4F4F4}{\color[HTML]{000000} 40 } & \cellcolor[HTML]{F4F4F4}{\color[HTML]{000000} 50 } & \cellcolor[HTML]{E6E6E6}{\color[HTML]{000000} 38 } &
\cellcolor[HTML]{E6E6E6}{\color[HTML]{000000} 35 } &
\cellcolor[HTML]{A9A9A9}{\color[HTML]{000000} 22 } & \cellcolor[HTML]{A9A9A9}{\color[HTML]{000000} 25 } & \cellcolor[HTML]{919191}{\color[HTML]{000000} 10 } & \cellcolor[HTML]{A9A9A9}{\color[HTML]{000000} 25 } \\

{\color[HTML]{000000} Xalan} & 
\cellcolor[HTML]{E6E6E6}{\color[HTML]{000000} 30 } & \cellcolor[HTML]{F4F4F4}{\color[HTML]{000000} 45 } & \cellcolor[HTML]{E6E6E6}{\color[HTML]{000000} 35 } & \cellcolor[HTML]{E6E6E6}{\color[HTML]{000000} 30 } & 
\cellcolor[HTML]{A9A9A9}{\color[HTML]{000000} 20 } &
\cellcolor[HTML]{A9A9A9}{\color[HTML]{000000} 25 } & \cellcolor[HTML]{919191}{\color[HTML]{000000} 10 } &
\cellcolor[HTML]{A9A9A9}{\color[HTML]{000000} 23 } & \cellcolor[HTML]{919191}{\color[HTML]{000000} 5 } \\

{\color[HTML]{000000} Ant} & 
\cellcolor[HTML]{919191}{\color[HTML]{000000} 18 } & \cellcolor[HTML]{A9A9A9}{\color[HTML]{000000} 20 } & \cellcolor[HTML]{A9A9A9}{\color[HTML]{000000} 20 } & \cellcolor[HTML]{919191}{\color[HTML]{000000} 15 } &
\cellcolor[HTML]{A9A9A9}{\color[HTML]{000000} 23 } &
\cellcolor[HTML]{F4F4F4}{\color[HTML]{000000} 50 } & \cellcolor[HTML]{A9A9A9}{\color[HTML]{000000} 22 } & \cellcolor[HTML]{A9A9A9}{\color[HTML]{000000} 22 } & \cellcolor[HTML]{A9A9A9}{\color[HTML]{000000} 20 } \\

{\color[HTML]{000000} Velocity} & 
\cellcolor[HTML]{E6E6E6}{\color[HTML]{000000} 34 } & \cellcolor[HTML]{F4F4F4}{\color[HTML]{000000} 50 } & \cellcolor[HTML]{F4F4F4}{\color[HTML]{000000} 45 } & \cellcolor[HTML]{E6E6E6}{\color[HTML]{000000} 35 } &
\cellcolor[HTML]{E6E6E6}{\color[HTML]{000000} 39 } &
\cellcolor[HTML]{A9A9A9}{\color[HTML]{000000} 20 } & \cellcolor[HTML]{919191}{\color[HTML]{000000} 10 } & \cellcolor[HTML]{919191}{\color[HTML]{000000} 10 } & \cellcolor[HTML]{919191}{\color[HTML]{000000} 10 } \\

{\color[HTML]{000000} Poi} & 
\cellcolor[HTML]{919191}{\color[HTML]{000000} 18 } & \cellcolor[HTML]{A9A9A9}{\color[HTML]{000000} 20 } & \cellcolor[HTML]{A9A9A9}{\color[HTML]{000000} 20 } & \cellcolor[HTML]{A9A9A9}{\color[HTML]{000000} 20 } &
\cellcolor[HTML]{A9A9A9}{\color[HTML]{000000} 20 } &
\cellcolor[HTML]{919191}{\color[HTML]{000000} 15 } & \cellcolor[HTML]{A9A9A9}{\color[HTML]{000000} 20 } & \cellcolor[HTML]{A9A9A9}{\color[HTML]{000000} 20 } & \cellcolor[HTML]{A9A9A9}{\color[HTML]{000000} 20 } \\

{\color[HTML]{000000} Synapse} & \cellcolor[HTML]{A9A9A9}{\color[HTML]{000000} 25 } & \cellcolor[HTML]{E6E6E6}{\color[HTML]{000000} 35 } & \cellcolor[HTML]{F4F4F4}{\color[HTML]{000000} 44 } & \cellcolor[HTML]{E6E6E6}{\color[HTML]{000000} 39 } &
\cellcolor[HTML]{E6E6E6}{\color[HTML]{000000} 34 } &
\cellcolor[HTML]{E6E6E6}{\color[HTML]{000000} 39 } & \cellcolor[HTML]{E6E6E6}{\color[HTML]{000000} 30 } & \cellcolor[HTML]{E6E6E6}{\color[HTML]{000000} 35 } & \cellcolor[HTML]{E6E6E6}{\color[HTML]{000000} 34 }\\
\hline
\end{tabular}
\caption{{\bf RQ2 results}:
IQR overlap scores: for the same median scores, smaller IQRs are better, marked in darker color.
\BLACK}
\label{tbl:iqr_overlap}
\end{table}

\subsection{RQ3: Is TimeLIME better at defect reduction?}

As discussed earlier, better plans in defect reduction field are believed to be those that are (a) easier to apply while (b) maintaining the effectiveness in reducing bugs. 
The first criterion has already been met.
As seen there, the plans made by TimeLIME
are much smaller, hence easier to apply, than the other
methods studied here. Also, as seen above, the plans
from TimeLIME correspond well to the known actions of developers.

The visualized result in Table \ref{tbl:weighted} shows that 3 variants of TimeLIME planners have obtained highest average $S_{scaled}$ scores in most of the projects (8 out of 9).  


The overall result is very clear: 
\begin{blockquote}
    \noindent
    \textbf{Answer 3}:   The   changes proposed by TimeLIME are associated with a much larger reduction in defects than classic LIME and other benchmark algorithms.
    \end{blockquote}
    
\section{Discussion}\label{discussion}
A potential major objection to all the above could be that the planning process, as we described so far, may be inefficient due to:
\bi
\item Developers may not be able to implement our plans;
\item Even if developers could implement the plans, they might  inadvertently make other changes that negate the improvements suggested in the plans.
\ei
In this section, we will discuss the practicality of our approach concerning these 2 issues, followed by another section of other, less pressing, threats to validity. To that end, we extended our measurement of similarity between the proposed plans and the actual actions. We further label each change in a single plan into one of the following 4 categories:
\bi
\item \textbf{T}rue \textbf{P}ositive: suggests   same change as seen later;
\item \textbf{T}rue \textbf{N}egative: suggests no change, and no change later;
\item \textbf{F}alse \textbf{P}ositive: suggests a change which is not seen later;
\item \textbf{F}alse \textbf{N}egative: suggests no change, but some other change is found later.
\ei

\noindent
We also calculate the precision and recall as defined in Table~\ref{tab:cm}:

\bi
\item \textbf{Precision = TP/(TP+FP)}: Among all changes proposed by a planner, how many of them are found in the next release? 
\item \textbf{Recall = TP/(TP+FN)}: For   changes found in the next release, how many  are the same as the planner's?
\ei

\begin{table}[t!]
\centering
     
\begin{tabular}{|r|r|r|r|r|r|r|r|r|r|}
\hline
{\color[HTML]{000000} } & {\color[HTML]{000000} \rotatebox{90}{Random} } & {\color[HTML]{000000} \rotatebox{90}{Oliveira}} & {\color[HTML]{000000} \rotatebox{90}{Shatnawi}} & {\color[HTML]{000000} \rotatebox{90}{Alves}} & {\color[HTML]{000000} \rotatebox{90}{XTREE}} &{\color[HTML]{000000} \rotatebox{90}{LIME}} & {\color[HTML]{000000} \rotatebox{90}{TimeLIME}} & {\color[HTML]{000000} \rotatebox{90}{TimeLIME\_MLP}} & {\color[HTML]{000000}  \rotatebox{90}{TimeLIME\_SVM}} \\
\hline
{\color[HTML]{000000} Jedit} & \cellcolor[HTML]{BFBFBF}{\color[HTML]{000000} 33 } &  \cellcolor[HTML]{BFBFBF}{\color[HTML]{000000} 34 } &
\cellcolor[HTML]{FFFFFF }{\color[HTML]{000000} 19 }& \cellcolor[HTML]{BFBFBF}{\color[HTML]{000000} 31 } &
\cellcolor[HTML]{BFBFBF}{\color[HTML]{000000} 40 } &
\cellcolor[HTML]{BFBFBF}{\color[HTML]{000000} 43 } & \cellcolor[HTML]{919191}{\color[HTML]{000000} 86 } & \cellcolor[HTML]{BFBFBF}{\color[HTML]{000000} 84 } & \cellcolor[HTML]{919191}{\color[HTML]{000000} 86 } \\
{\color[HTML]{000000} Camel1} & \cellcolor[HTML]{BFBFBF}{\color[HTML]{000000} 45 } & \cellcolor[HTML]{BFBFBF}{\color[HTML]{000000} 55 } & \cellcolor[HTML]{FFFFFF }{\color[HTML]{000000} 16 } &  \cellcolor[HTML]{BFBFBF}{\color[HTML]{000000} 50 } &
\cellcolor[HTML]{BFBFBF}{\color[HTML]{000000} 62 } &   
\cellcolor[HTML]{BFBFBF}{\color[HTML]{000000} 77 } & \cellcolor[HTML]{BFBFBF}{\color[HTML]{000000} 84 } & \cellcolor[HTML]{BFBFBF}{\color[HTML]{000000} 85 } & \cellcolor[HTML]{919191}{\color[HTML]{000000} 86 } \\
{\color[HTML]{000000} Camel2} & \cellcolor[HTML]{BFBFBF}{\color[HTML]{000000} 35 } & \cellcolor[HTML]{BFBFBF}{\color[HTML]{000000} 42 } & \cellcolor[HTML]{FFFFFF }{\color[HTML]{000000} 13 } &  \cellcolor[HTML]{BFBFBF}{\color[HTML]{000000} 38 } &
\cellcolor[HTML]{BFBFBF}{\color[HTML]{000000} 35 } &   
\cellcolor[HTML]{BFBFBF}{\color[HTML]{000000} 63 } & \cellcolor[HTML]{BFBFBF}{\color[HTML]{000000} 65 } & \cellcolor[HTML]{919191}{\color[HTML]{000000} 74 } & \cellcolor[HTML]{BFBFBF}{\color[HTML]{000000} 65 } \\
{\color[HTML]{000000} Log4j} & \cellcolor[HTML]{BFBFBF}{\color[HTML]{000000} 33 } & \cellcolor[HTML]{BFBFBF}{\color[HTML]{000000} 44 } & \cellcolor[HTML]{FFFFFF}{\color[HTML]{000000} 23 } &  \cellcolor[HTML]{BFBFBF}{\color[HTML]{000000} 39 } &
\cellcolor[HTML]{BFBFBF}{\color[HTML]{000000} 42 } &
\cellcolor[HTML]{BFBFBF}{\color[HTML]{000000} 43 } & \cellcolor[HTML]{919191}{\color[HTML]{000000} 78 } & \cellcolor[HTML]{BFBFBF}{\color[HTML]{000000} 77 } & \cellcolor[HTML]{BFBFBF}{\color[HTML]{000000} 75 } \\

{\color[HTML]{000000} Xalan} & 
\cellcolor[HTML]{FFFFFF }{\color[HTML]{000000} 60 } & \cellcolor[HTML]{BFBFBF}{\color[HTML]{000000} 81 } & \cellcolor[HTML]{FFFFFF}{\color[HTML]{000000} 60 } &  \cellcolor[HTML]{BFBFBF}{\color[HTML]{000000} 74 } & 
\cellcolor[HTML]{BFBFBF}{\color[HTML]{000000} 67 } &
\cellcolor[HTML]{BFBFBF}{\color[HTML]{000000} 85 } & \cellcolor[HTML]{919191}{\color[HTML]{000000} 96 } & \cellcolor[HTML]{BFBFBF}{\color[HTML]{000000} 95 } & \cellcolor[HTML]{BFBFBF}{\color[HTML]{000000} 95 } \\

{\color[HTML]{000000} Ant} & 
\cellcolor[HTML]{BFBFBF}{\color[HTML]{000000} 49 } & \cellcolor[HTML]{BFBFBF}{\color[HTML]{000000} 58 } & \cellcolor[HTML]{FFFFFF }{\color[HTML]{000000} 18 } &  \cellcolor[HTML]{BFBFBF}{\color[HTML]{000000} 47 } &
\cellcolor[HTML]{BFBFBF}{\color[HTML]{000000} 62 } &
\cellcolor[HTML]{919191}{\color[HTML]{000000} 99 } & \cellcolor[HTML]{BFBFBF}{\color[HTML]{000000} 94 } & \cellcolor[HTML]{BFBFBF}{\color[HTML]{000000} 97 } & \cellcolor[HTML]{BFBFBF}{\color[HTML]{000000} 92 } \\

{\color[HTML]{000000} Velocity} & 
\cellcolor[HTML]{BFBFBF}{\color[HTML]{000000} 50 } & \cellcolor[HTML]{BFBFBF}{\color[HTML]{000000} 58 } & \cellcolor[HTML]{FFFFFF }{\color[HTML]{000000} 18 } &  \cellcolor[HTML]{BFBFBF}{\color[HTML]{000000} 47 } &
\cellcolor[HTML]{BFBFBF}{\color[HTML]{000000} 52 } &
\cellcolor[HTML]{BFBFBF}{\color[HTML]{000000} 48 } & \cellcolor[HTML]{919191}{\color[HTML]{000000} 91 } & \cellcolor[HTML]{919191}{\color[HTML]{000000} 91 } & \cellcolor[HTML]{919191}{\color[HTML]{000000} 91 } \\

{\color[HTML]{000000} Poi} & 
\cellcolor[HTML]{BFBFBF}{\color[HTML]{000000} 38 } & \cellcolor[HTML]{BFBFBF}{\color[HTML]{000000} 50 } & \cellcolor[HTML]{FFFFFF}{\color[HTML]{000000} 0 } &  \cellcolor[HTML]{BFBFBF}{\color[HTML]{000000} 50 } &
\cellcolor[HTML]{BFBFBF}{\color[HTML]{000000} 54 } &
\cellcolor[HTML]{BFBFBF}{\color[HTML]{000000} 43 } & \cellcolor[HTML]{BFBFBF}{\color[HTML]{000000} 75 } & \cellcolor[HTML]{919191}{\color[HTML]{000000} 77 } & \cellcolor[HTML]{BFBFBF}{\color[HTML]{000000} 74 } \\

{\color[HTML]{000000} Synapse} & \cellcolor[HTML]{BFBFBF}{\color[HTML]{000000} 37 } & \cellcolor[HTML]{BFBFBF}{\color[HTML]{000000} 41 } & \cellcolor[HTML]{FFFFFF }{\color[HTML]{000000} 1 } &  \cellcolor[HTML]{BFBFBF}{\color[HTML]{000000} 36 } &
\cellcolor[HTML]{BFBFBF}{\color[HTML]{000000} 41 } &
\cellcolor[HTML]{BFBFBF}{\color[HTML]{000000} 60 } & \cellcolor[HTML]{919191}{\color[HTML]{000000} 71 } & \cellcolor[HTML]{BFBFBF}{\color[HTML]{000000} 65 } & \cellcolor[HTML]{BFBFBF}{\color[HTML]{000000} 66 }\\
\hline
\end{tabular}
\caption{{\bf RQ3 results}: 
Improvement percentage per project: the higher the better. Best and worst planner in each project are marked in dark and light respectively.  
\\
}
\label{tbl:weighted}
\end{table}
\begin{table}[t!]
\centering
\begin{tabular}{|l|l|l|}
\hline
 &
  Actual: change &
  Actual: no / different change \\ \hline
TimeLIME: change &
  TP &
  FP \\ \hline
TimeLIME: don't change &
  FN &
  TN\\ \hline
\end{tabular}
\caption{ Each change/no-change proposed in a plan will be categorized into one of the 4 kinds according to the actual value in the {\em most recent release}. Example can be found in Table. \ref{tab:overlap} \BLACK}
\label{tab:cm}
\end{table}
\noindent
Unanticipated changes that are not recommended by a planner are marked as \textbf{FN}. Plans proposed by TimeLIME but not happened in the next release are marked as \textbf{FP}. A higher \textbf{precision} means more of TimeLIME's plans are undertaken, and a higher \textbf{recall} means there are fewer unanticipated changes in the next release. If it is TimeLIME rather than unanticipated changes that should be credited for the reduced defects, then ideally the defect-reducing plans should be associated with a high precision, and among those plans, most of them should also have a high recall.

First, to evaluate if developers are capable of implementing changes proposed by our plans, we measure the \textbf{precision} rates of plans from different algorithms. The result from Table \ref{tab:precision} shows that both LIME and TimeLIME planner obtain the highest scores. This is a supportive evidence indicating
that developers, as seen in the subsequent release, were capable of implementing most of the changes proposed by our plans.

Secondly, to answer the question that whether or not developers may inadvertently make other changes while following plans proposed by planners, we measure the \textbf{recall} rates of plans. A low recall means that there exist more unanticipated changes. Therefore, it is more questionable that whether the plan or the unforeseen changes should take credit for the effect of defect reduction. As seen in Table. \ref{tab:recall}, TimeLIME and LIME still have the highest recall rates among all algorithms, which makes the performance of the planners more convincing since it is revealed here that when developers are making changes proposed by LIME or TimeLIME, they are less likely to deploy other changes that are not mentioned.

\begin{table}[t!]
\centering
\begin{adjustbox}{max width=\linewidth}
\begin{tabular}{|l|l|l|l|l|l|l|l|}
\hline
& Xtree & Shat & Oliv & Alves & Random & TimeLIME    & LIME \\ \hline
Jedit    & 38    & 0    & 7    & 5     & 7      & 65 & 58   \\ 
Camel1   & 80    & 0    & 8    & 7     & 7      & 90 & 73   \\ 
Camel2   & 22    & 0    & 3    & 6     & 5      & 83 & 66   \\ 
Log4j    & 23    & 1    & 10   & 8     & 12     & 81 & 70   \\ 
Xalan    & 34    & 1    & 1    & 2     & 1      & 81 & 82   \\ 
Ant      & 21    & 0    & 10   & 8     & 1      & 50 & 49   \\ 
Velocity & 19    & 0    & 2    & 3     & 13     & 88 & 85   \\ 
Poi      & 45    & 1    & 4    & 18    & 4      & 61 & 76   \\ 
Synapse  & 39    & 0    & 18   & 18    & 4      & 70 & 66   \\ \hline
AVG & 35.67 & 0.33 & 7.00           & 8.33 & 6.00          & 74.33 & 69.44 \\ \hline
STD & 17.99 & 0.47 & 5.01 & 5.52 & 4.03 & 12.79 & 10.69 \\ \hline
Rank     & 2     & 4    & 3    & 3     & 3      & 1  & 1    \\ \hline
\end{tabular}
\end{adjustbox}
\caption{ The precision rate (in percentage) of a plan measures how many changes proposed by the plan are found in the subsequent release. The rank is generated using the Scott-Knot test. A higher rank is better.\BLACK}
\label{tab:precision}
\end{table}

\begin{table}[t!]
\centering
\begin{adjustbox}{max width=\linewidth}

\begin{tabular}{|l|l|l|l|l|l|l|l|}
\hline
  & Xtree & Shat & Oliv & Alves & Random & TimeLIME    & LIME  \\ \hline
Jedit    & 23    & 0    & 3    & 5     & 2      & 59    & 64    \\ 
Camel1   & 29    & 0    & 2    & 3     & 3      & 59    & 52    \\ 
Camel2   & 18    & 0    & 1    & 1     & 2      & 62    & 64    \\ 
Log4j    & 5     & 1    & 8    & 6     & 6      & 58    & 72    \\ 
Xalan    & 61    & 0    & 1    & 2     & 1      & 85    & 81    \\ 
Ant      & 11    & 0    & 2    & 6     & 5      & 33    & 31    \\ 
Velocity & 27    & 0    & 1    & 3     & 2      & 73    & 77    \\ 
Poi      & 30    & 0    & 0    & 6     & 2      & 60    & 56    \\ 
Synapse  & 23    & 0    & 2    & 2     & 1      & 46    & 52    \\ \hline
AVG      & 25.22 & 0.11 & 2.22 & 3.78  & 2.67   & 59.44 & 61.00 \\ \hline
STD      & 14.88 & 0.31 & 2.20 & 1.87  & 1.63   & 13.85 & 14.46 \\ \hline
Rank     & 2     & 4    & 3    & 3     & 3      & 1     & 1     \\ \hline
\end{tabular}
\end{adjustbox}
\caption{ The recall rate (in percentage) of a plan measures out of all actual changes how many of them get proposed by the plan. \BLACK}
\label{tab:recall}
\end{table}

In summary, by mining the historical releases, the evaluational analysis here shows sufficient evidence that TimeLIME is of greater practicality compared to other algorithms. Furthermore, We also believe that more studies could be done to explore and expand the value of current evaluation process.
\BLACK

\section{Threats to validity}\label{threat}
Due to the complexity of the experiment designed in this case study, there are many factors that can threaten the validity of these results.
\subsection{Learner Bias} 
This paper selects Random Forest classifier as the black-box classifier because prior research has shown that Random Forest classifier is ranked as one of the top models among all 32 classifiers used in defect prediction~\cite{ghotra2015revisiting}. However, the preeminent predictive power of Random Forest classifier does not ensure that explanations derived from it are preeminent code refactoring plans as well. Other methods from the top rank may be more suitable in the problem of explanation generation while we haven't explored more.

\subsection{Instrument Bias}
Various approaches are proposed for explainable AI. Although LIME is one of the widely cited and well-known tools, it  other tools might be  suitable for solving SE problems, which can make solutions from LIME sub-optimal. Hence, to verify if adding in SE knowledge can always improve AI tools, we need to make a comprehensive exploration that includes more explanation generation methods.
\subsection{Hyper-parameter Tuning}
Past researches have shown how hyper-parameter optimization can boost the performance of a classifier used in defect prediction. Since in this paper we concentrate on the modification of the explainer instead of the learner, we used a simple grid search to find the optimal parameter setting. It can be possible that the current setting is sub-optimal and by using the actually optimal settings we might receive different experiment result.

\section{Related work}\label{related}
Much research urges that interpretability should become an important factor in assessing analytical models in software engineering because software developers expect the model to provide understandable suggestions that can be actually achieved in real-world practice \cite{menzies2013software,lewis2013does,dam2018explainable}.

Recently at TSE'20, Jiarpakdee et al. modified LIME using hyper parameter optimization techniques, and assessed its performance in defect prediction via output stability \cite{jiarpakdee2020empirical}. The result has shown that explanations generated from their method are not only more stable among re-generations, but also understandable to software developers. 
The major difference is that: 
\begin{itemize}[leftmargin=16pt,topsep=4pt]

\item 
Jiarpakdee et al. assess the viability of applying model-agnostic techniques (such as LIME) in defect prediction whereas this paper assesses the practical effectiveness of LIME in re-organizing a project
\item 
Jiarpakdee et al. explore possible means to improve the explanation generation procedure where as
this paper explores methods to refine LIME's results into more actionable and effective plans for defect reduction.   

\end{itemize}

\section{Future work}\label{futurework}
For future work, we need to take action to retire the above threats
to validity. 
\subsection{More Learners}
More black-box learners should be used in the experiment to construct a more comprehensive comparison. Although the limited sample amount of defect prediction datasets has ruled out many deep learning models such as Neural Network due to the overhead, there are still many other models, including but not limited to Random Subspace Sampling and Sequential Minimal Optimization, applicable for this experiment.
\subsection{More Explainers}
As described above, LIME is a representative member in the family of local surrogate interpretation models. Other local explanation generation methods that apply tree-structure extraction or association rule mining or so on should also be introduced in the discussion.
\subsection{More Data}
We would like to collect not only more SE projects of defection prediction data but also more releases of a single project. This can facilitate the further exploration on the accountability of our historical data analysis. According to the $K$-test, we validate plans on the more current release of the 3 releases.
Because of that, we would prefer the files in the validation release is more similar to the proposed plans, no matter they have more or fewer bugs, so that our evaluation on the plans can be more accountable. 
Sometimes when the file in the validation release is not similar to the proposed plan, we wonder what would the file be like if there is another release with a more similar file. Could it be possible that more releases can provide us more accurate and robust evaluation conclusions?

\subsection{Better Data}
Our current data collection methods
use all available training data.
This means our model is learning from
the track records of all developers.
One potential issue with that approach is that  the  defect  reduction  performed  by  the  development  teams may  not  be  the  best.  For  example,  say  that  in  release 2, one  team  removed  5  defects  with  respect  to  release 1.
Potentially, another team with
more experience could have removed
ten  defects  instead.  Yet the above study learns from both teams even though one may be more effective than the other.
  
This is a  fruitful
area for useful future work.
For example, in future work we could 
mine GitHub repositories looking for   ``tourists''; i.e. people that usually work in classes A,B,C,.., but then (occasionally)
make changes to sections of the code that
they are not so familiar with (class D,E, etc).  Potentially,
if we remove the ``tourist'' data and only learn from activities of more experienced developers,
we might be able to build better plans using the better-refined training data. 
\BLACK

\subsection{More Measurements: Multi-objective Optimization}
In this paper, we introduced $K$-test as a framework to conduction quality evaluation on changes proposed by different planners. However, although the current framework does provide us with insightful findings, we still believe that more measurements need to be brought in to construct a more comprehensive evaluation process. As shown in \S\ref{discussion}, planners make changes of different sizes, which makes it harder to examine their effectiveness since they are from different levels of precision: suppose 2 planners both made a change that overlaps with a defect-reducing action, the planner with a more precise/smaller change interval should probably be considered better than the other one. To address this problem, the future work could import another fitness score function that relates the precision to the effectiveness of the plan. That is to say, the task of planning defect reduction could be regarded as a multi-objective optimization problem, where a planner might have several goals (effectiveness, precision, feasibility, etc.) to chase after at the same time. 

\section{Conclusion}\label{conclusion}
This paper has assessed
the following {\em TimeLIME tactic} for generating
defect reduction plans:
\begin{quote}
{\em 
When reasoning about changes
to a project, it is best
to use changes seen in the historical record of that project.}
\end{quote}
Using this tactic, we find plans that:
\begin{itemize}
    \item \emph{Are succinct}: In terms of the average size of recommended plans. The TimeLIME generally generates smaller plans than the classical LIME and RandomWalk in every project. The plans are also equivalently succinct compared to other benchmark methods in this paper. Smaller plans are preferred to larger plan since the latter can be faster to apply.
    \item \emph{Better resemble developers' actions}: In terms of the overlap between the proposed plans and the developer actions in the upcoming release, plans proposed by TimeLIME better match what developers actually do.
    \item \emph{Are better at reducing defects}: In terms of the scaled weighted scores $S_{scaled}$ that indicate the overall net gain received per project. TimeLIME gets the highest score among all planners in 8 out of 9 trials. 
    (while the classical LIME wins in only 1 project).
\end{itemize}
Our results are a cautionary tale to the SE community.
SE researchers need to be  more careful about using off-the-shelf AI tools, without first tuning them with SE knowledge.
Specifically:
\begin{quote}
{\em
It is 
unwise to throw standard AI tools at SE problems without first
  considering how those tools might be customized for SE applications.
  }
\end{quote}
We say that since,
as shown here, (a)~such customization is not a complex thing to do and (b)~the customized system can have   dramatically better performance.

\section*{Acknowledgements}
This work was partially funded by 
a research grant from the National Science Foundation (CCF \#1703487) and the Laboratory for Analytical Sciences, North Carolina State University.


\bibliographystyle{IEEEtran}
\bibliography{main} 
\begin{IEEEbiography}{Kewen Peng} 
is a second year Ph.D. student in Computer Science at North Carolina State University.  
His research interests include using and refining artificial intelligence methods to solve problems in software engineering.
\end{IEEEbiography}
\begin{IEEEbiography}[{\includegraphics[width=1in,clip,keepaspectratio]{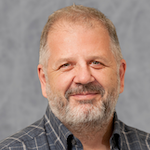}}]{Tim Menzies} (IEEE Fellow)
is a Professor in CS at North Carolina State University.  
His research interests include software engineering (SE), data mining, artificial intelligence, search-based SE, and open access science. \url{http://menzies.us}
\end{IEEEbiography}

\clearpage
\end{document}